# A quiet STEVE disturbs navigation satellites' signals in the Antarctic

Daria Kotova[1*], Luca Spogli[2], Yaqi Jin[1], and Wojciech Miloch[1]

[1]*Department of Physics, University of Oslo, Oslo, Norway*
[2]*Istituto Nazionale di Geofisica e Vulcanologia, Rome, Italy*

## Abstract

Strong Thermal Emission Velocity Enhancement (STEVE) is a narrow optical phenomenon that occurs equatorward of the auroral oval and is associated with intense subauroral plasma flows and thermospheric heating. Although these conditions can produce plasma irregularities, direct evidence of STEVE effects on radio wave propagation during geomagnetically quiet conditions has so far not been observed. Here we report for the first time a STEVE event observed over Antarctica during quiet geomagnetic conditions and show that it produced measurable fluctuations in the Global Navigation Satellite System (GNSS) signals. Using coordinated optical observations and high-resolution (50 Hz) GNSS scintillation measurements from two Antarctic stations, we identify enhanced phase and amplitude scintillation coincident with intersections between GNSS signal paths and the STEVE arc. The observations indicate the presence of plasma irregularities and suggest substantial temporal variations in the apparent altitude of the optical structure (vertical motion within the 130–270 km range). We find that even a relatively weak STEVE event can affect the GNSS signal propagation in the absence of major geomagnetic disturbances, extending previous studies that associated such effects primarily with storms and intense auroral activity. This finding expands the understanding of STEVE's geophysical impact and highlights potential vulnerabilities in satellite-based navigation systems during seemingly benign space weather conditions. The fact that this STEVE event occurred just off the Antarctic coastline, where scientific expeditions and seagoing vessels rely heavily on precise positioning indicates that STEVE-related plasma structuring should be considered in assessments of the GNSS performance under otherwise quiet geomagnetic conditions in subauroral regions.

## Introduction

Over the past decade, scientists have been attracted to an unusual optical phenomenon, later named STEVE [MacDonald et al., 2018]. STEVE is an acronym for a Strong Thermal Emission Velocity Enhancement and is represented by a thin line of violet-white glow or purple/mauve arc appearing slightly equatorward of the auroral glow, often accompanied by distant green bands of ascending lines called a picket fence [Nishimura et al., 2023a]. In recent years, we have investigated many aspects about this beautiful and rather rare subauroral phenomenon [Gallardo-Lacourt et al., 2021; Nishimura et al., 2023a]. Outcomes have shown that STEVE is an optical manifestation of the presence of intense subauroral ion drift (SAID) in the region and is closely related to it [Archer et al., 2019a; Mishin & Streltsov, 2023; Nishimura et al., 2023a]. The main features of STEVE are the same as for SAID: observation of density depletion (midlatitude density trough), an enhancement of electron temperature, and the presence of a fast flow in thin subauroral regions. For most of the events considered in the literature, the ion velocity exceeds 3 km/s [Archer et al., 2019a]. The study by Zhang et al. [2024] shows that even higher drift velocities (up to 20 km / s) occur within STEVE.

The conducted studies of the height of STEVE manifestation point to a range of 130-270 km [Archer et al., 2019b] which is mainly located in the F-region of the ionosphere. STEVE is identified by its airglow continuum within the 400-700 nm range [Gillies et al., 2019s; Liang et al., 2019] emissions. Its spectrum highlights two primary components: the OI red-line emissions at 630.0 and 636.4 nm, and a broad enhancement of a continuous spectrum spanning approximately 400-730 nm [Gallardo-Lacourt et al., 2021]. Despite its proximity to the aurora, STEVE lacks particle precipitation (electrons and ions) and is associated with low conductance in the subauroral region [Gallardo-Lacourt et al., 2018a; Nishimura et al., 2023a]. The study by Liang et al. [2019] first presented a clear discrepancy between STEVE and stable auroral red-line.

Statistical studies [Gallardo-Lacourt et al., 2018b; Archer et al., 2019a] provided important information on the time and frequency of STEVE occurrence. Thus, the main duration of STEVE is about an hour and the time of occurrence is from 22 to 02 magnetic local time within a range of 59°-63° magnetic latitudes. STEVE has a rather narrow optical structure with a 20-50 km width, and the longitudinal duration is about 2000 km. It also manifests itself with a seasonal dependence, with the most frequent observations occurring during the equinoxes and the least frequent at solstices (based on statistics for the northern hemisphere). An anti-correlation with solar activity has also been reported. However, as Gallardo-Lacourt et al. [2018b] highlighted, an expansion of the statistics is required for a more confident statement of the solar cycle, seasonal and annual frequency of STEVE observations. Despite this, it is worth noting that these statistics are consistent with the seasonal dependence of SAID observations, which was built on a large base of Defense Meteorological Satellite Program (DMSP) satellite observations for 25 years [He et al., 2014]. Not every SAID event is accompanied by STEVE, but each STEVE is associated with unusually intense SAID events.

Although SAID events are most commonly observed during the recovery phase of a substorm, STEVE is not typically associated with high geomagnetic activity [Gallardo-Lacourt et al., 2018b]. Based on 28 events STEVE occurs during positive Bx and negative By and Bz components of the Interplanetary Magnetic Field (IMF) and about an hour after the substorm onset toward the end of an extended substorm expansion phase (and Kp about 4). The recent paper by Gallardo-Lacourt et al. [2024] showed different mechanisms for the STEVE formation in non-storm and non-substorm conditions. To this day, STEVE remains a hot scientific topic and leaves many open scientific challenges [Nishimura et al., 2023a]. One such question is raised in the current article: How does STEVE affect the propagation of trans-ionospheric radio waves? While a previous study has linked STEVE-driven ionospheric irregularities to global navigation satellite systems (GNSS) disruptions during geomagnetically active periods [Chen et al., 2024b], this work presents the first evidence of STEVE-induced scintillations under quiet geomagnetic conditions (Kp ~$2^-$). These findings challenge the assumption that such subauroral phenomena require significant magnetospheric disturbances to affect trans-ionospheric signals. Hence, our paper details the disturbances induced on GNSS when STEVE and the picket fence were observed with optical instruments in Antarctica, Troll station under geomagnetically quiet conditions. Using data from high-resolution scintillation receivers we show the influence of STEVE on the GNSS signal during quiet geomagnetic conditions for the first time.

## Main

### Methods

The STEVE event on May 9, 2019, was examined here in detail, when, according to the all-sky camera (Figure 1), STEVE was captured appearing as a purple arc and accompanied by a picket fence. We used data from the Keo Sentry3 all-sky imager (ASI) that was installed in January 2019 at the Norwegian Research Station Troll in Dronning Maud Land, Antarctica. The geographic and geomagnetic coordinates of the station are, respectively, 72.0016°S, 2.5254°E and 62.7476°S MLAT, 47.8047°E MLONG. Magnetic coordinates were obtained in the altitude-adjusted corrected geomagnetic (AACGM) coordinates system [Shepherd, 2014] for the selected event, 22:33 Magnetic Local Time (MLT) for the station corresponding to midnight at UT. ASI operates with 20 seconds step three wavelengths of 557.7 nm (green line) and 630.0 nm (red line) for respectively atomic oxygen O(1S) and O(1D) auroral and airglow emission lines, and 427.8 nm (blue line) for $N_2^+$ (1N) emissions in the Earth's atmosphere. Thus, the time resolution on each wavelength is 1 minute.

The Troll station (TRL) is also equipped with the NovAtel GPStation-6 multi-constellation and multi-frequency receiver for scintillation monitoring. From this device we use signals from GPS (L1CA and L5), GLONASS (L1CA and L2P), and Galileo (E1, E5a, and E5b) satellite systems, delivering raw observational data such as 50 Hz phase and amplitude measurements. In addition to that, we used data from the Septentrio PolaRx5S scintillation monitoring receiver (Bougard et al., 2011) installed at the SANAE IV station (SNA). The station is located relatively close to the Troll station, approximately 186 km to the northwest. The geographic and geomagnetic coordinates of SNA are, respectively, 71.6735°S, 2.8409°W and 61.8877°S MLAT, 44.9625°E MLONG, 22:22 MLT. Data from the PolaRx5s receivers at SNA are part of the data collections available in the electronic Space Weather upper atmosphere (eSWua, eswua.ingv.it) data portal (Upper atmosphere physics and radiopropagation Working Group, 2020). From the SNA scintillation receiver, we use signals with a 50 Hz rate from the GPS (L1CA, L2C, and L5) and Galileo (E1, E5a, and E5b) satellite systems. The reasonable noise levels for PolaRx5S receiver were calculated based on data during the quiet conditions over the mid-latitude station in Ushuaia (ARG), showing noise levels for amplitude scintillation index S4 = 0.015 and for phase scintillation index $\sigma_\phi$ = 0.015 rad. Therefore, the value for both indices 0.03 (as written in the manual for $\sigma_\phi$) is a good excess approximation and is a quite good approximation for the GPStation6 receiver too.

We also use data from two fluxgate magnetometers at the SANAE IV and Neumayer III (VNA) stations. The geographic and geomagnetic coordinates of the last one are, respectively, 70.683°S, 8.282°W and 60.5888°S MLAT, 42.31835°E MLONG, 22:11 MLT. Sensors of both magnetometers have HDZ orientation where H is the annual mean value of horizontal intensity, D is declination and Z is vertical down. The mean baseline is subtracted from values. All three stations are located on the edge of a quiet auroral oval providing a unique opportunity to study the subauroral region in detail.

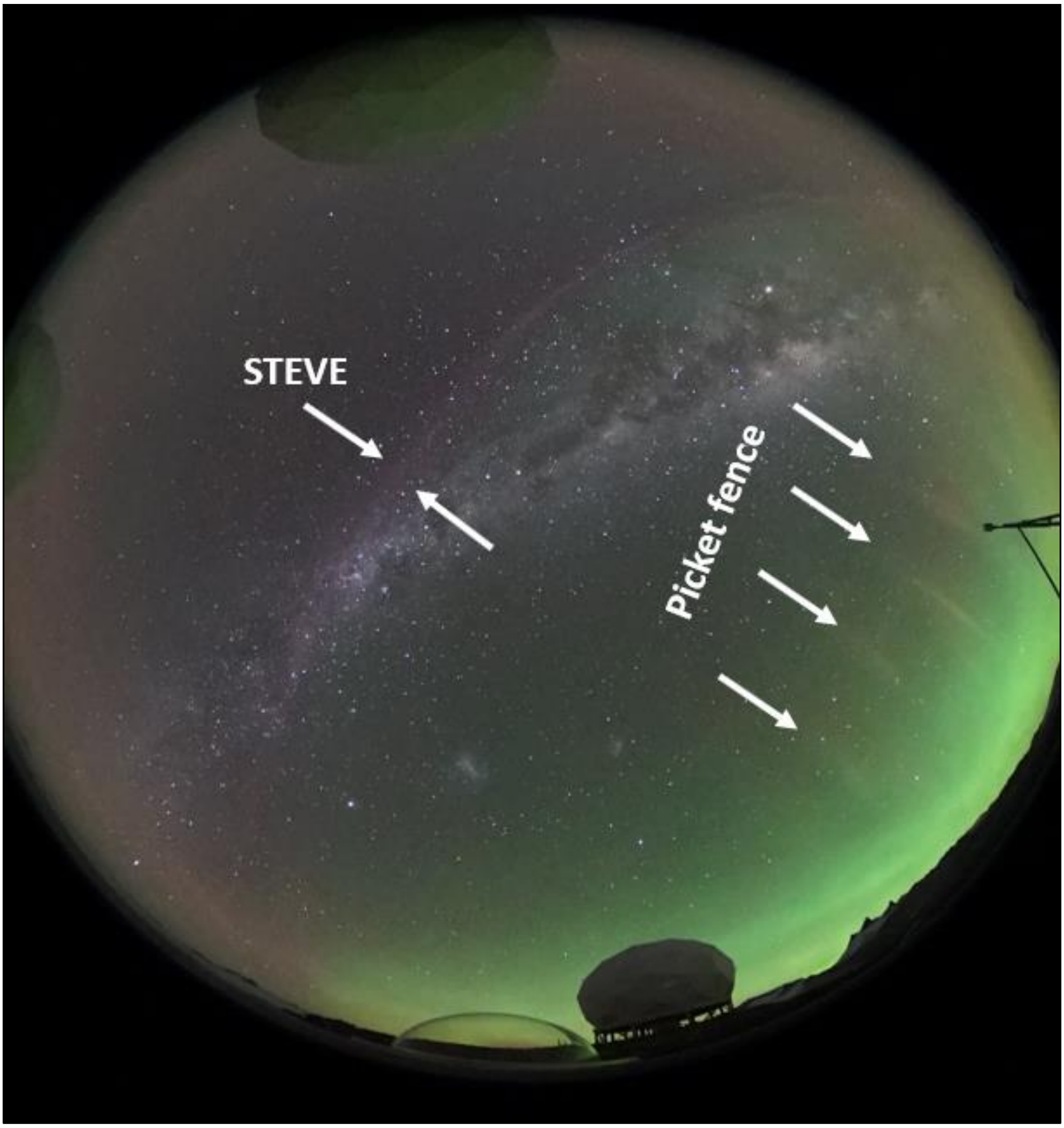

**Figure 1.** Image from the all-sky colour camera (Sony α7SII) installed at the Troll station, Antarctica. The snapshot captures STEVE and the picket fence over the station on 2019-05-09 at 23:42:50 UT.

The low-Earth orbit (LEO) satellite coverage over Dronning Maud Land during the STEVE event was sparse, and no clean in situ data were available to characterise the ionospheric conditions or drift velocities (more information provided in the Supplementary Information, Figs. S7-S8). The ASI orientation was calibrated by the stars, and each pixel was recalculated to the heights of the STEVE observation range (150 km, 170 km, 230 km, and 250 km) [Archer et al., 2019b]. We use a mask of 10 degrees above the horizon for ASI. In the work, only the 630 nm red-line emission filter data are mainly presented, since, as was also noted in the literature [Gillies et al., 2019; Liang et al., 2019; Gallardo-Lacourt et al., 2021], STEVE is visually better represented by this emission line (see Figure S9 in SI). The coordinates of the subionospheric points (ionospheric pierce point, IPP) used for the projection of the satellite-receiver beams to one ionospheric height were recalculated to the same grid as ASI. To display the IPP on all-sky images, we used 1-minute scintillation indices and a 20° elevation mask provided by receivers. In addition to it, we used 50 Hz dual-frequency phase measurements to calculate 1-second scintillation indices, and we first focused on high-resolution measurements to see in detail the effect of STEVE on satellite signals. The use of high-rate scintillation indices has proven to be highly effective in identifying small-scale irregularities within the auroral oval (Enengl et al., 2023; Nishimura et al., 2023), whereas traditional 1-minute values often fail to detect them (Spogli et al., 2009). A 6th-order Butterworth high-pass filter with a cutoff frequency of 0.1 Hz was used to find the detrended raw carrier phase and standard equations to find the phase [van Dierendick et al., 1993] and amplitude [Briggs & Parkin, 1963] scintillation indices from the high-resolution data. To characterize the geophysical conditions, we use the upstream IMF data from the OMNI dataset time-shifted to the Earth's bow shock [King & Papitashvili, 2005]. The geomagnetic indices: the SYM-H index and auroral electrojet (AE) indices from World Data Center Tokyo, Kp index from GFZ https://kp.gfz-potsdam.de/ and data from magnetometers on SNA and VNA stations.

We also calculated the ionosphere-free linear combination (IFLC) when two frequencies were available to gain further insight into the scintillation generation mechanisms [Carrano et al., 2013; McCaffrey & Jayachandran, 2019; Ghobadi et al. 2020; Spogli et al., 2021; Zheng et al., 2022; Enengl et al., 2023, Miloch et al., 2024]. The IFLC was derived using the following formula:

$$\Phi_{IFLC} = \frac{\Phi_{L1} f_{L1}^2 - \Phi_{L2} f_{L2}^2}{f_{L1}^2 - f_{L2}^2} \quad (1)$$

where $f_{L1}$ and $f_{L2}$ are the carrier frequencies of the two signals, and $\Phi_{L1}$ and $\Phi_{L2}$ are the corresponding carrier phase measurements. This combination effectively removes the first-order ionospheric delay due to the refractive effects of the ionosphere, allowing us to isolate the higher-order effects and better understand the underlying scintillation processes as diffractive variations that are usually a sign of the presence of small-scale irregularities.

## The STEVE and geomagnetic conditions

If the initial analysis had been based on 1Hz scintillation receiver and magnetometer data, which show no disturbances (see Supported Information, Figure S1), such a day would have been excluded from the analysis. The advantage of visual analysis of the All-Sky Imager (ASI, see more information in the Methods section) data allowed us to identify this event (note that there are only a few overwintering teams and no civilian photographs in Antarctica that could help, as is often the case in STEVE-related studies). The geomagnetic conditions of the selected event can be described as quiet space weather conditions (see Figure 2). The Kp index was low ($2^-$), indicating minimal geomagnetic activity on a global scale. The SYM-H index reached around -7 nT and does not show any signature of stormy conditions, indicating a weak ring current activity. The AE index reached 400 nT around 08 UT and 250 nT during STEVE manifestation. Values of 400 nT and 250 nT indicate some moderate auroral activity but no significant disturbances in the auroral zone. The average magnitude of the IMF was about 4.46 nT with the Bx-component predominantly positive, and By- and Bz- components negative during STEVE. This condition repeats the discovered IMF conditions in [Gallardo-Lacourt et al., 2018b; Gallardo-Lacourt et al., 2024] that likely contributed to the occurrence of STEVE. According to the ground-based magnetometers at the SANAE IV (Figure 2e) and Neumayer III stations (Figure 2f), no significant bursts in magnetic field components were observed in the STEVE observation region. These conditions generally mean that space weather is calm with a period of relative stability in Earth's magnetosphere and ionosphere. However, STEVE was observed for about one hour with ASI from 23:18 to 00:18 UT, which agrees with the usual observation time and STEVE duration [Gallardo-Lacourt et al., 2018b].

STEVE is formed on the equatorial boundary of the diffuse aurora (see Fig. 3 and Video in SI) within the 62°S and 65°S magnetic coordinates. STEVE is visible over several MLTs and limited by the ASI field-of-view, where the eastern part is more related to night conditions and the western to evening ones. Therefore, on the night side, STEVE descends closer to the equator. It can be noted that STEVE is more extended along geomagnetic latitudes than along geographic ones. First, STEVE is formed poleward to the considered stations and rapidly begins to move equatorward from 23:18 to 23:41 UT. Around 23:37 UT a stable STEVE arc is formed south of the Troll station. After 23:41 UT there are successive minor shifts toward the pole, toward the equator and back to the pole until STEVE finally dissolves in the night sky around 00:18 UT.

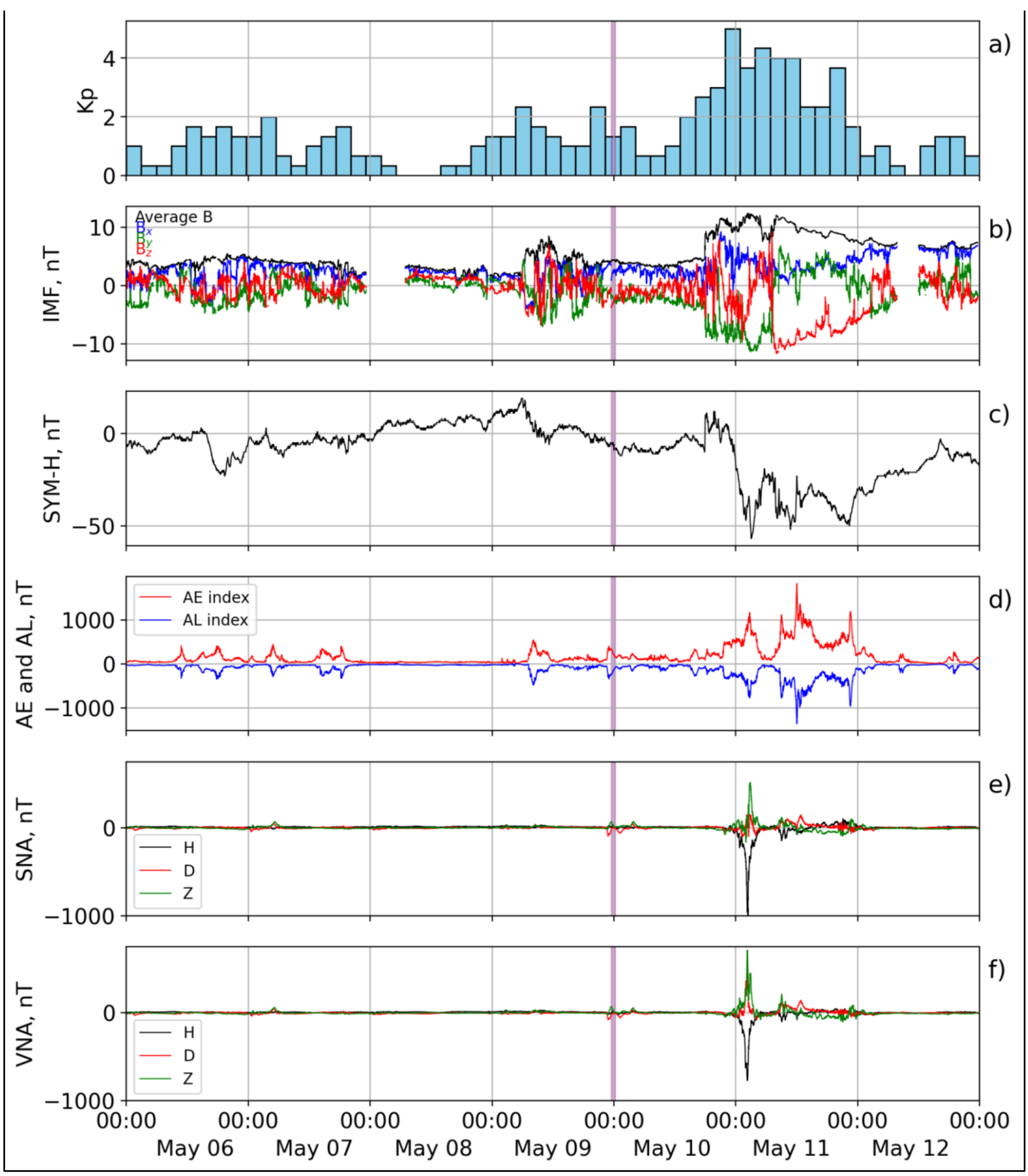

**Figure 2.** Geomagnetic activity and interplanetary magnetic field (IMF) conditions during May 06-12, 2019. a) three-hourly Kp index. b) the interplanetary magnetic field components, c) the SYM-H index, and d) the Auroral Electrojet Indices AE/AL from the OMNI database. e) Magnetometer data from SANAE IV (SNA) and f) from the Neumayer III station (VNA) in HDZ sensor orientation where H presents horizontal intensity. The purple fill presents the time interval of STEVE observation.

**Results**

We demonstrate for the first time STEVE-triggered effects on high-resolution GNSS scintillation data. During the STEVE lifetime, about 19 satellites have crossed or passed closely to its optical appearance (Figure 3). Figures 4-8 show some examples of these occurrences.

The results were obtained based on 50Hz data from both receivers. Only high-elevation observations are considered to exclude fluctuations caused by multipath reflections from terrestrial objects. Panel (a) displays the pixel intensity at the satellite projection point on the ASI. For a better understanding of the flyby dynamics, we recommend watching the video in the supplementary information. The panels below show 1-second values of phase $\sigma_\phi$ and amplitude S4 scintillation indices for different frequencies. The figure is completed by showing the ionosphere-free linear combination (IFLC, see more information in the Methods section) behaviour. The elevation angle is also reported along with S4, to highlight possible effects due to low elevation observations. Satellite numbers here and further in the text will be discussed following the coding adopted for Figure 3 (PRN+0 correspond to GPS, PRN+40 correspond to GLONASS, and PRN+70 correspond to Galileo). In addition, the abbreviation G is used for GPS, E for Galileo, and R for GLONASS.

Since the GNSS receiver and the ASI at the Troll station are co-located, the relative position of the data sets (ionospheric pierce points, or IPPs, and STEVE on ASI data) remains consistent regardless of the projection altitude. This means that only one height needs to be considered for the projections for the receiver at Troll (TRL), simplifying the analysis (Figs 7 and 8). Different projection heights are visually checked for the SANAE IV station (SNA is code fore receiver) to find the best agreement between the data sets (more examples can be found in Supporting Information). The presence of variation in the scintillation indices for the Troll receiver during the spatial passage of the satellite's IPP through STEVE indicates the importance of checking ASI frames for different altitudes. By visually confirming the best correlation, we can assume the height of the STEVE flow. Thus, the most suitable projection heights were selected: 230 km for the E26 (Fig. 4) and G08 (Fig. 6) satellites, and 250 km for E18 (Fig. 5). These heights correspond to the upper limit of STEVE [Archer et al., 2019b].

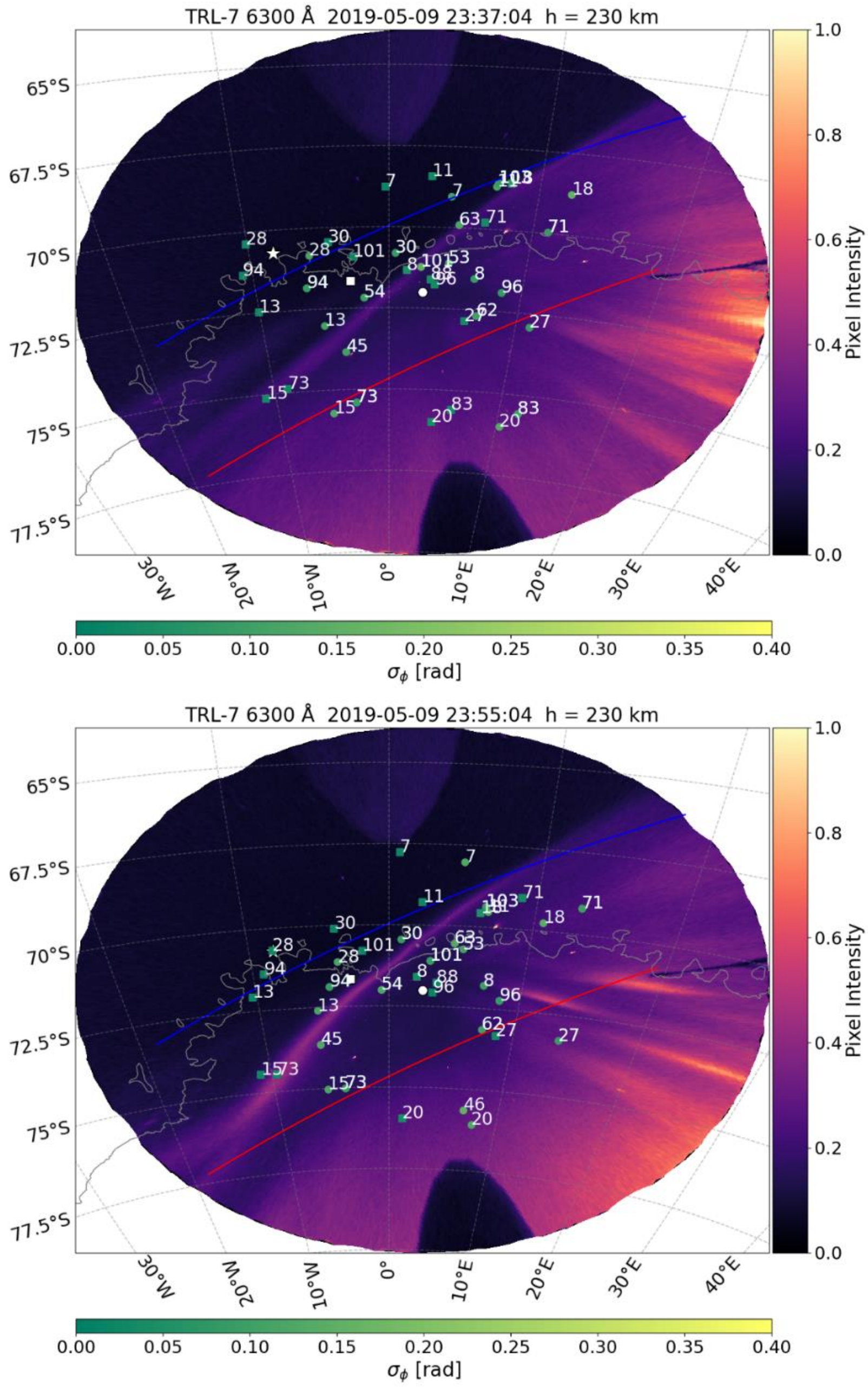

**Figure 3.** The 630 nm images from the all-sky imager were projected at an altitude of 230 km together with satellite's IPPs. The white star presents the position of the Neumayer III station. The white circle and square show the locations of the Troll (TRL) and SANAE IV (SNA) stations respectively, and the corresponding shapes show the calculated IPP for the satellites for the selected altitude, as recorded by receivers on the ground. Only satellites with elevation angles above 20° are presented here. The colour bars show the intensity of the 1-minute phase scintillations index (below) and the pixel intensity (to the right of the images). The following code is used to distinguish different GNSS: PRN+0 correspond to GPS (G), PRN+40 correspond to GLONASS (R), and PRN+70 correspond to Galileo (E). The blue and red lines show the AACGM geomagnetic isolines of 62°S and 65°S.

All five satellites (SNA E26, SNA E18, SNA G08, TRL E31, TRL R13) in question experience variations in $\sigma_\phi$ on all frequencies from 23:17 UT to 23:52 UT. The observed disturbances are not significant and only exceed the noise level. However, these variations can be two times or more greater than the noise level and stand out against the background of the quiet signal (see also more information in the Methods section). For the SNA receiver the noise level is around 0.03 for both $\sigma_\phi$ and S4 and is represented in the figures as a dashed red line. Small variations in the amplitude scintillation index are observed on satellites SNA G08 (Fig. 6), TRL E31 (Fig. 7) and TRL R13 (Fig. 8). It is noteworthy that the S4 variations on G08 are present only on the L1CA frequency and begin a bit earlier than the $\sigma_\phi$ and glow intensity peaks. An interesting case is the one of the two satellites: SNA E18 and SNA E26, which are near each other when crossing STEVE. Satellite E18 enters the STEVE front almost perpendicularly, which is reflected in a clear peak in the phase scintillation index at 23:31 UT (on both frequencies E1 and E5a, see Fig. 5), while satellite E26 passes more tangentially. For this reason, variations in the phase measurements on satellite E26 are above the noise level from 23:18 UT (especially at E5a) whereas on satellite E18 we observe a gradual increase in $\sigma_\phi$. However, $\sigma_\phi$ from E26 reaches two clear peaks at 23:31 UT and 23:37 UT (Fig. 4). It is noteworthy that the peak at 23:37 UT agrees well with the projection of STEVE and IPP to an altitude of 170 km (see SI Fig. S2). Both satellites later enter the zone of discrete auroras. Still, it is clearly seen that for both satellites no changes in intensity, which are related to auroral particle precipitation, lead to an increase in $\sigma_\phi$ above the noise level. The selection of different projection heights of STEVE shows how dynamic the structure is on the vertical scale.

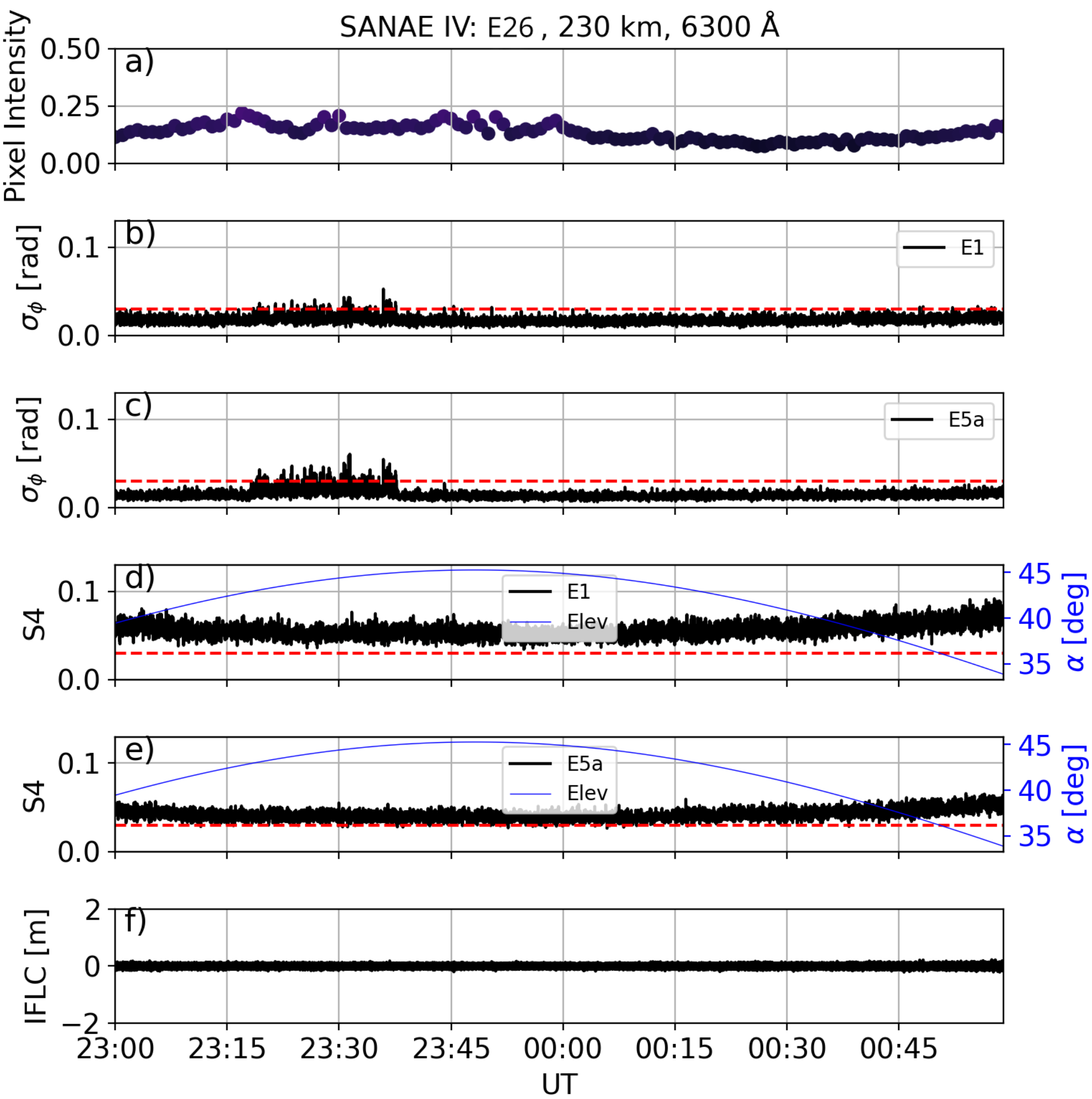

**Figure 4.** (a) Pixel intensity at the respective IPP position of the Galileo satellite PRN E26 (96), 1-second $\sigma_\phi$ for (b) E1 and (c) E5a frequencies recorded by the Septentrio PolaRx5S receiver, 1-second S4 for (d) E1 and (e) E5a signals plotted together with the elevation angle in blue (axis in the right), (f) ionospheric free linear combination (IFLC). The red dotted line indicates the noise level of the receiver.

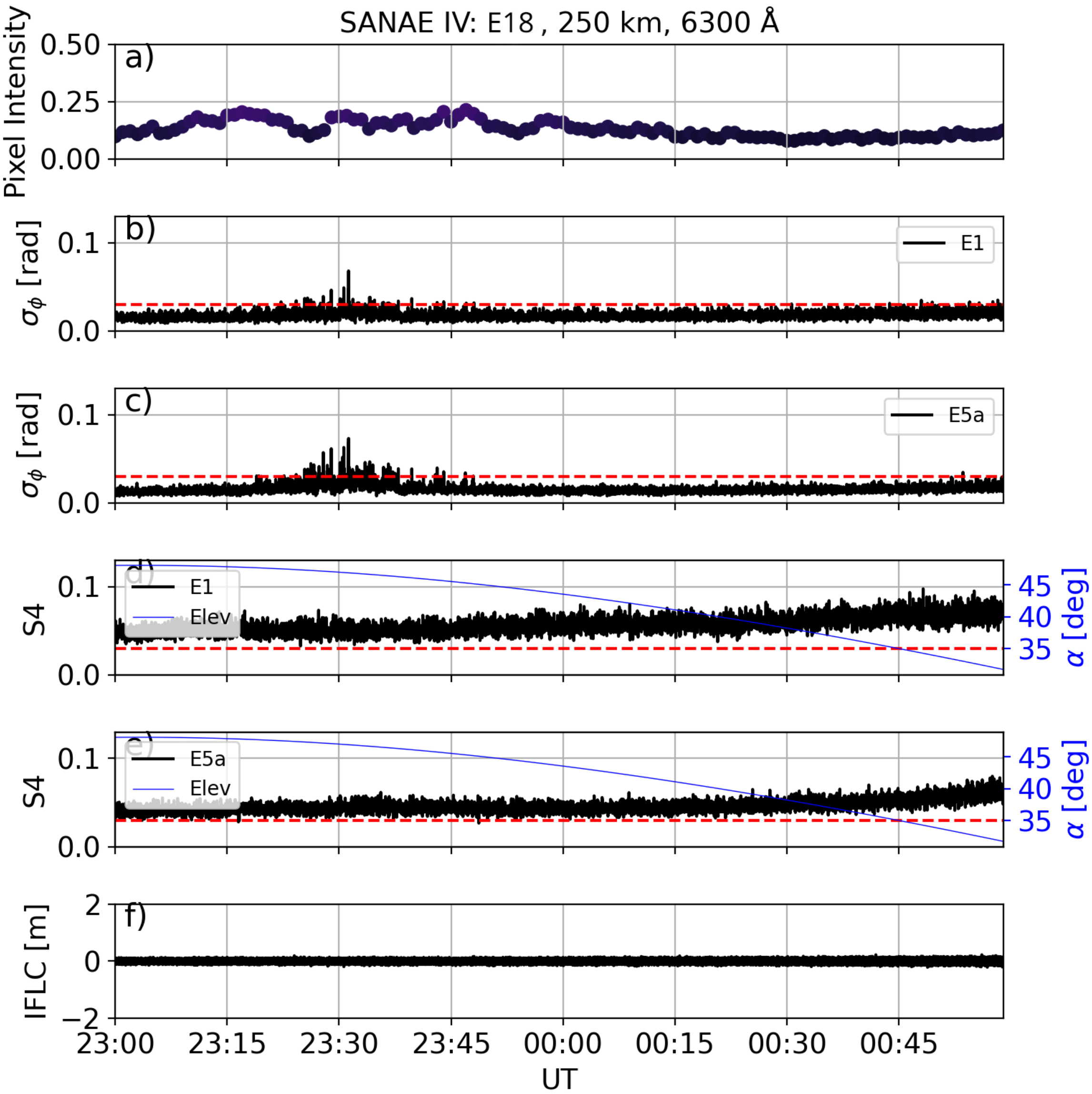

**Figure 5.** Same as Figure 4 but for Galileo PRN E18 (88).

Satellite SNA G08 (Fig. 6) is caught in the dynamic shift of the STEVE front, first encountering the leading edge of the equatorward moving front and then the poleward moving front, which effectively places the satellite in the region of this phenomenon. This is evidenced by a strong correlation between the increase in phase fluctuations, indicated by $\sigma_\phi$ (panels b-d), and the position of STEVE (panel a). The maximum of $\sigma_\phi$ for three frequencies was observed at 23:47 UT when the satellite exited the STEVE region. Notably, this phase scintillation index peak aligns better with the projection at 170 km (see SI Fig. S3), suggesting that the front edge of STEVE is located at higher altitudes of 230-250 km, while the trailing edge (or the dynamic progression of the phenomenon over time) descends to 170-200 km. Interestingly, although amplitude scintillations were more pronounced on the L2C and L5 signals [Salles et al., 2021], in this instance, the effect of STEVE is observed only at the higher signal frequency L1CA.

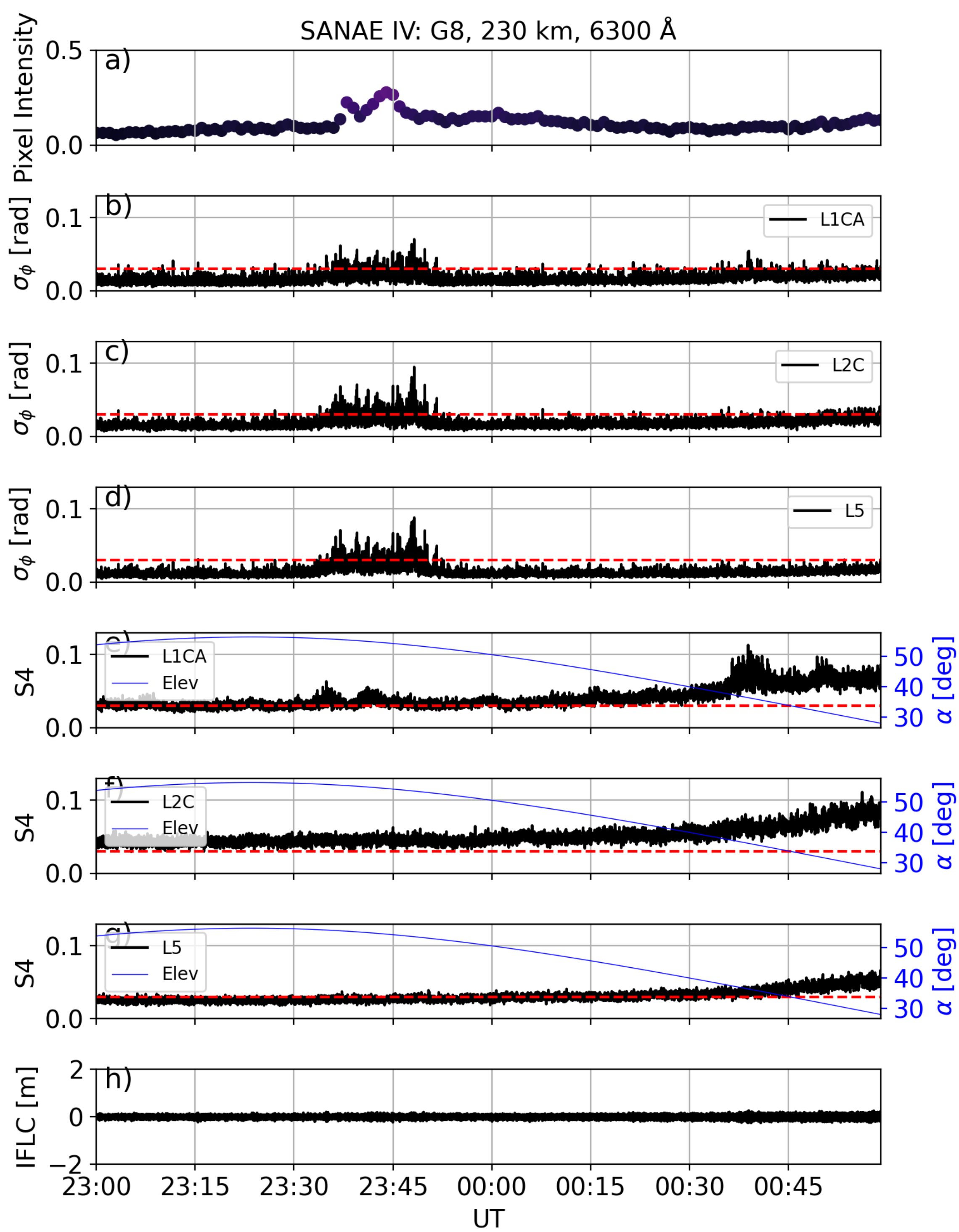

**Figure 6.** As Figure 4 but for GPS PRN G08 (8) with scintillation indexes $\sigma_\phi$ and S4 calculated for L1CA (b and e), L2C (c and f), and L5 (d and g) signals.

The next pair of satellites (Figs. 7 and 8) is characterized by an even higher elevation angle than the previously considered ones. The results from the TRL E31 satellite (Fig. 7) align perfectly with theoretical expectations, showing weak amplitude variations (S4) at the lower frequencies of the E2 and E5 signals (panels f and g), while such variations are absent on E1 (panel e). The significant increase in the phase scintillation index during passage through the STEVE front is also higher for the E2 and E5 signals than for E1. As the satellite moves parallel to STEVE, the signal does not experience any further significant fluctuations in both amplitude and phase measurements. It is worth noting that the phase measurements from the receiver at Troll are noisier compared to those from the receiver at SANAE IV. This is due to the different types of scintillation receivers, with the PolaRx5s receiver exhibiting slightly lower noise levels.

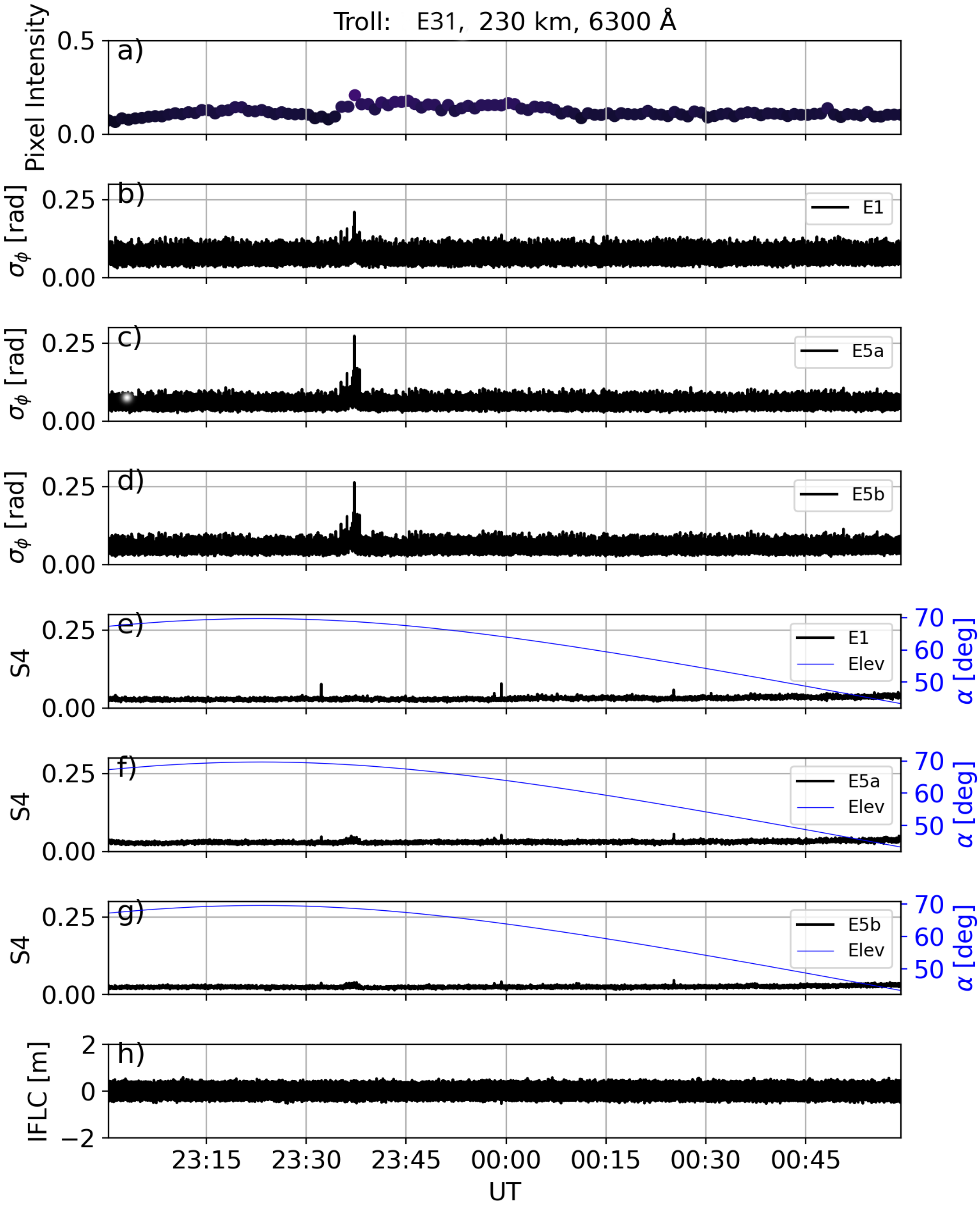

**Figure 7.** Same as Figure 4 but for Galileo PRN E31 (101) recorded by the NovAtel GPStation-6 receiver at the Troll station. Scintillation indexes $\sigma_\phi$ and S4 were calculated for E1 (b and e), E5a (c and f), and E5b (d and g) signals.

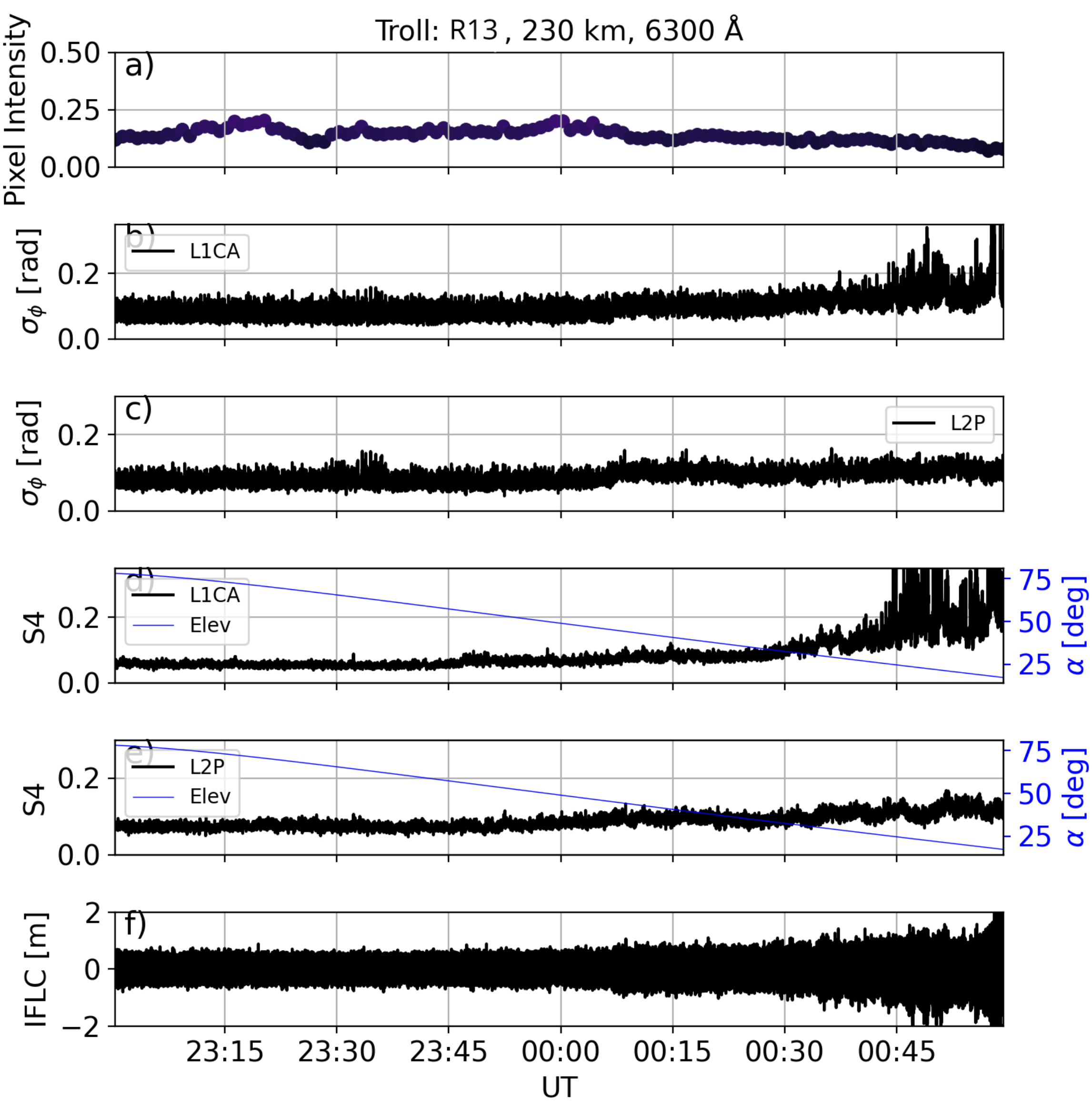

**Figure 8.** Same as Figure 7 but for GLONASS PRN R13 (53) for L1CA and L2P signals.

The last example of satellite TRL R13, although not distinguished by significant changes in the measurements of both scintillation indices, shows small variations in the signal during the passage of the STEVE region from 23:30 UT to 23:37 UT. This satellite also moves tangentially, smoothly entering and exiting STEVE (similar results show R14 Fig. S4 but a bit later 23:37 UT to 23:49 UT). Variations in both $\sigma_\phi$ and S4 indices after 00:30 UT at the L1CA frequency (panels b and d) spatially coincide with the region of previous presence of STEVE. As the satellite descends closer to the horizon, it moves along the STEVE boundary and likely experiences variations from density irregularities remaining in this region. A decrease in signal power makes it more sensitive to changes in the propagation environment. Similar variations are also visible in Fig. 6f and e, but in this case, the G08 satellite enters the aurora region. However, as seen in Fig. 8f, the variations in both indices for the R13 satellite are associated with diffraction effects in the ionosphere, which may indirectly indicate the presence of small-scale electron density structures in the signal passage zone. In the remaining cases considered, the increase in scintillation indices is associated with refractive effects, as the IFLC (the last panels in Figs. 4-8) does not show visible changes for all the cases considered.

# Discussion

In this article, we analyse high-resolution 50 Hz data from two scintillation receivers located in Dronning Maud Land, Antarctica (also known as Queen Maud Land). Both receivers are located on the equatorial boundary of the quiet auroral oval (at Troll and SANAE IV), which provides a unique opportunity to study in detail the influence of subauroral phenomena such as STEVE. The co-located all-sky camera and imager allow a spatial examination of the STEVE evolution, as well as make assumptions about its height and evolution of height of STEVE during the event, and to assess the impact on GNSS signals. Geomagnetic indices of the considered event on May 9, 2019, describe reasonably quiet conditions (Fig. 2). Ground-based measurements from magnetometers at SANAE IV and Neumayer III also show no significant local variations during the event, confirming at a local level the overall quiet geomagnetic conditions. However, STEVE is also manifested in such conditions (Fig. 3). Although STEVE is usually associated with SAID events and is described in the literature as an optical manifestation of a strong/intense SAID event [Archer et al., 2019a; Mishin & Streltsov, 2023; Nishimura et al., 2023a], there are also indications that STEVE was developed in less active geomagnetic condition [Gallardo-Lacourt et al., 2024]. The question arises whether STEVE events can be considered as a quiet ionosphere. And what is a quiet ionosphere? Our study shows that even a weak STEVE can produce measurable fluctuations in phase and amplitude measurements. It is quite possible that more serious STEVE events can lead to more serious consequences of trans-ionospheric signals. This was recently addressed in a study by Chen et al. [2024b] under geomagnetically disturbed conditions, which, in synergy with our work, helps completing the initial picture of STEVE's effects on GNSS. This can be a very critical point for the correct positioning for ship traffic and aviation in the region close to the Antarctic coastline, where there are both tourist and science operations. Previous work by Kotova et al. [2025] has shown that the presence of a SAID event can lead to a serious deterioration of the received GNSS signal up to the point of loss of satellite tracking. However, the effect of strong STEVE requires further study, consideration of more serious geomagnetic events, and a statistical analysis of its effects on GNSS under both quiet and disturbed conditions.

All the GNSS satellites considered show the effect of STEVE at elevation angles greater than 40°, which allows us to exclude the effects of multipath. Moreover, satellites passing the discrete aurora zone (or "picket fence") do not experience variations discussed above (see SI Fig. S5 as an example). The question remains why not all satellites crossing the region of manifestation of STEVE experience signal fluctuations. Thus, observational geometry plays a role. Indeed, the reported examples clearly show that the satellites crossing STEVE more perpendicularly presents stronger STEVE-related fluctuations than those tangentially. This can also be related to stronger gradients in case of a perpendicular intersection. When the satellite crosses tangentially, it gradually enters STEVE with westward fast flow and drifted fine-scale structures [Nishimura et al., 2023b] inside. STEVE itself moves also along the N-S direction. The perpendicularly crossing satellites are also significantly affected by the movement of the STEVE front. Additionally, most examples of crossings were in correspondence with the eastern part of STEVE, featured by the presence of the picket fence, as observed with the green emission. This is likely the time and location where STEVE reached its maximum development. Most of the satellites from the west were following closely STEVE with no obvious signal fluctuations. It is worth remembering that this is a weak STEVE developed in quiet geomagnetic conditions. Another point may be the different types and sensitivity of scintillation receivers used in the study. In the data from the Troll station, disturbances are less visible because the phase measurements have higher noise levels. De Paula et al., [2021] show that raw data for GPStation-6 presented very large abnormal upward fluctuations (incursions).

Chen et al. [2024a] using triangulation of data from two cameras with all-sky fish-eye lenses, show a change in STEVE altitude over time with a more stable at 160 km and short-term excursion up to 200 km within a span of approximately 4 minutes. Previous estimates by Chu et al. [2019] show that along the purple arc STEVE's altitude did not change significantly. In our case, using different projections of the IPP and ASI altitude, it was possible to observe a very dynamic change in the altitude of STEVE from 230-250 km to 170-200 km in 10 minutes (see Fig. 6 for the SNA G08 satellite). The example of the same satellite shows that the trailing edge of STEVE influences the signal more strongly than the leading (or frontal, due to the initial dynamics of STEVE development towards the equator). The work by Nishimura et al. [2023a] shows that the leading edge drifted westward faster than the picket fence that connected to the trailing edge of STEVE and extended along similar magnetic field lines [Archer et al., 2019]. Liang et al., [2019] show that the higher elevated STEVE corresponds to 'pinkish' emission and the lower elevated STEVE to 'whitish'. This would indicate that the largest scintillations should have been observed at the leading edge of the STEVE. Therefore, it is difficult to conclude whether this is a special case, or an effect of the dynamic structure. To confirm our results, additional datasets are required. As noted in the Methods section, unfortunately, LEO observations from Swarm and DMSP are not available for this event. Further research is needed to answer this question. The event reported here is a rare event presenting STEVE during quiet conditions and it highlights the importance of using GNSS-related information to better understand STEVE and related phenomena.

Additionally, our ASI observations of STEVE do not reveal visually fine-scale structures. However, Nishimura et al. [2023b] suggest the existence of small-scale structures, even those smaller than 1 km, with a potential to cascade to smaller scales through turbulence. We observe variation in the S4 index that can be a sign of a small-scale irregularities but mostly for L1 frequency at one satellite G08 (Fig. 6). In fact, amplitude scintillation is generally expected to decrease with increasing frequency, meaning lower-frequency signals experience stronger amplitude fluctuations due to the power-law relationship (see, e.g., Song et al., 2021 and references therein). Consequently, L1 C/A (1575.42 MHz) should exhibit less intense amplitude scintillations than L2C (1227.60 MHz) and L5 (1176.45 MHz). We are confident that this observed feature is not due to instrumental effects, as both scintillation receivers feature advanced signal processing techniques that ensure comparable tracking robustness across all GNSS bands. The very low level of S4 suggests a possible selectivity of smaller-scale irregularities. Specifically, assuming an altitude $h$ of 230 km for the STEVE phenomenon and an overhead propagation observational geometry, the Fresnel scales $d_F=(2\lambda h)^{0.5}$ for the three considered GNSS frequencies are: 296 m for L1 C/A, 335 m for L2C, and 342 m for L5 [Kintner et al., 2007]. This may indicate that only the smaller-scale irregularities are contributing to the observed weak scintillation patterns. In other words, in the found low-scintillation conditions due to STEVE irregularities, smaller irregularities can exist without larger ones [Wernik et al., 2003]. Additionally, this selectiveness in scale sizes can also be due to the ray path crosses those spatial gradients, the rate of crossing those gradients, creates a temporal change in the signals received. This temporal change can have similar characteristics to diffraction-caused scintillation.

Simultaneous analysis of the IFLC shows that, in all cases, the $\sigma_\phi$ fluctuations are associated with refraction effects. The presence of a strong gradient in electron density and temperature, combined with strong flow, will inevitably create conditions conducive to plasma instabilities. Several ideas for the mechanisms of instability generation have been raised in the literature: gradient drift instability (GDI) [Rathod et al., 2021a], temperature gradient instability (TGI) [Eltrass et al., 2014] and ionospheric feedback instability (IFI) [Mishin & Streltsov, 2023]. In addition, the presence of flow share with high-speed ion drift (because STEVE appears during SAID) can create conditions favourable for Kelvin-Helmholtz instability (KHI), leading to the development of wave structures and turbulence. The SuperDARN convection maps (presented in SI, Fig. S6) show that the Heppner-Maynard boundary was located a bit poleward of the Troll station. The simulation results of GDI by Rathod et al. [2021a] indicate that turbulent energy cascades to smaller scales, suggesting that GNSS scales could potentially be reached. Using high-resolution data from NorSat-1, Sinevich et al. [2021, 2022, 2023] demonstrated small-scale irregularities within SAID. Rathod et al. [2021b] showed through model results that KHI dominates above a certain altitude (around 700 km), while GDI dominates below (around 500 km). The authors also noted that different dominating instabilities are observed by satellites and ground-based systems in subauroral polarization streams. In our study, the satellite's path goes through the ionosphere to the receiver on the ground, and dominating instability effects will depend on the geometry of this path, as discussed in Kotova et al. [2025]. Eltrass et al. [2016] investigated the cascading processes of TGI and GDI in the midlatitude ionosphere (in context of subauroral region), revealing that TGI is the primary generation mechanism during quiet times, while both instabilities can occur during disturbed times. Model simulations show wave cascading of TGI from kilometer scales down to decameter scales. IFI can lead to the formation of small structures [Jia & Streltsov, 2014] and it has been recognized as a primary mechanism for generating small-scale field-aligned currents (FAC) in the auroral and subauroral regions [Streltsov & Mishin, 2018], as well as the picket fence structure associated with weak upward FAC [Nishimura et al., 2019]. Given the quiet geomagnetic conditions and the relative altitude of the co-located observations of STEVE (approximately 230 km), along with the observed variations in phase scintillations, we can assume that TGI and IFI can be the primary instabilities causing GNSS signals disturbances and leading to scintillation. However, we lack sufficient data presenting the exact positions and sizes of irregularities, making it difficult to prove and draw definitive conclusions about which specific instabilities influence GNSS signal propagation in our case.

## Conclusions

The novelty of this work is twofold. On the one hand, we present the first evidence of STEVE's impact on GNSS signals during quiet geomagnetic conditions on May 9, 2019, that we observe over Antarctica. On the other hand, we discuss and present the temporal evolution of potential altitude changes of STEVE. Using ground-based instruments in Dronning Maud Land, Antarctica, we present the second investigation of STEVE in the Southern Hemisphere (the first was made by Martinis et al., 2022) and the first for Antarctica. By analyzing data from the all-sky imager and high-resolution 50 Hz data from two scintillation receivers, we observe that STEVE can cause increased values of both 1-second phase ($\sigma_\phi$) and amplitude (S4) scintillation indices, exceeding the receiver noise level by two or more times. According to the all-sky imager, STEVE was visible for about one hour from 23:18 UT to 00:18 UT, accompanied by the characteristic green picket fence. The observed increases of the 1-second $\sigma_\phi$ and S4 indices align well with the passage of satellite projections (IPP) through the STEVE zone. We also demonstrate that satellites crossing the STEVE front perpendicularly exhibited more pronounced increases in the high-rate scintillation indices, while those passing tangentially appears to be not or little affected. Considering various heights of the IPP and ASI projections, we conclude that STEVE is a highly dynamic structure on the vertical scale, with height changes from 230-250 km to 170-200 km over 10 minutes. The analysis of IFLC indicate that the variations in phase and amplitude scintillation indices associated with STEVE are characterized by refractive effects in the ionosphere. The increased S4 index values suggest the presence of small-scale plasma structures with dimensions of several hundred meters inducing diffractive effects on the GNSS signals.

In conclusion, the findings of this study underscore the potential impact of even weak STEVE events on the GNSS signal integrity, leading to fluctuations in phase and amplitude measurements. This highlights the critical importance of understanding and monitoring STEVE events that appear in the subauroral regions, especially in regions like the Antarctic coastline, where accurate positioning is vital for scientific research traffic, ships, and civil aviation. Previous research by Kotova et al. [2025] has demonstrated that SAID events can severely degrade GNSS signals, sometimes resulting in the loss of satellite tracking. Therefore, further investigation into the effects of strong STEVE events, particularly during more intense geomagnetic conditions, is essential to mitigate potential disruptions to trans-ionospheric signals and ensure reliable navigation and communication systems. Understanding STEVE's origin and behaviour can provide insights into the complex interactions in the subauroral regions between the Earth's magnetosphere, ionosphere, and thermosphere.

## Data Availability

Magnetometer data from the SANAE IV research station is made available through the South African National Space Agency (SANSA) http://www.sansa.org.za. Data from the Neumayer III station can be obtained from the INTERMAGNET database (www.intermagnet.org). The Troll station datasets analysed during the current study are available in the zenodo repository, https://doi.org/10.5281/zenodo.21997672. Data from the PolaRx5s receivers at SANAE IV are part of the data collections available in the electronic Space Weather upper atmosphere (eSWua, eswua.ingv.it) data portal managed by INGV.

## Acknowledgements

DK, YJ, WM acknowledge funding from the European Research Council (ERC) under the European Union's Horizon 2020 research and innovation programme (ERC Consolidator Grant agreement No. 866357, POLAR-4DSpace). Troll Ionosphere Observatory at the Troll research station in Antarctica has been supported by the Research Council of Norway grant numbers 267408 and 322466. Authors thank Norwegian Polar Institute for technical and logistic support in operating Troll Ionospheric Observatory. DK acknowledge Dr. Lasse B. N. Clausen for helping with calibration of ASI for Troll. Authors acknowledge the use of magnetometer data from SANAE IV research station provided by the South African National Space Agency (SANSA) http://www.sansa.org.za. The ionospheric scintillation monitoring receiver in in SANAE IV was deployed in the framework of the DemoGRAPE project funded by Italian National Antarctic Program (PNRA, Progetto di Ricerca 2013/C3.01) and it is run under the Memorandum of Understanding between Istituto Nazionale di Geofisica e Vulcanologia (INGV) and SANSA (Delibera CdA INGV 297/2016). LS expresses his gratitude to Emanuele Pica, Carlo Marcocci, Rayan Imam, and Lucilla Alfonsi of INGV for the help and support with raw data from the SANAE IV receiver. Authors acknowledge the use of magnetometer data collected at Neumayer III Station and thank Alfred Wegener Institute (AWI) and GFZ Helmholtz Centre for Geosciences for supporting its operation and INTERMAGNET for promoting high standards of magnetic observatory practice (www.intermagnet.org).

## Author information

**Authors and Affiliations**

**Department of Physics, University of Oslo, Oslo, Norway**
Daria Kotova, Yaqi Jin, & Wojciech Miloch

**Istituto Nazionale di Geofisica e Vulcanologia, Rome, Italy**
Luca Spogli

**Contributions**

DK, LS, YJ, and WM conceived the idea of the study. LS provided SANAE IV data curation and scripts for work with it. DK conducted the data analysis and prepared figures. DK wrote the text of the manuscript. All authors discussed the results and worked on the final version of the manuscript and its revision.

**Corresponding author**

Correspondence to Daria Kotova (daria.kotova@fys.uio.no).

## Conflict of interests

The authors declare no competing interests.

# Supporting Information

## A quiet STEVE disturbs navigation satellites' signals in the Antarctic

Daria Kotova*, Luca Spogli, Yaqi Jin, and Wojciech Miloch
Corresponding author: Daria Kotova (dariakot@fys.uio.no)

### Contents of this file

Supplementary Figures S1-S9, videos: 09052019_150km_sig_phi.mp4, 09052019_170km_sig_phi.mp4, 09052019_230km_sig_phi.mp4, and 09052019_250km_sig_phi.mp4

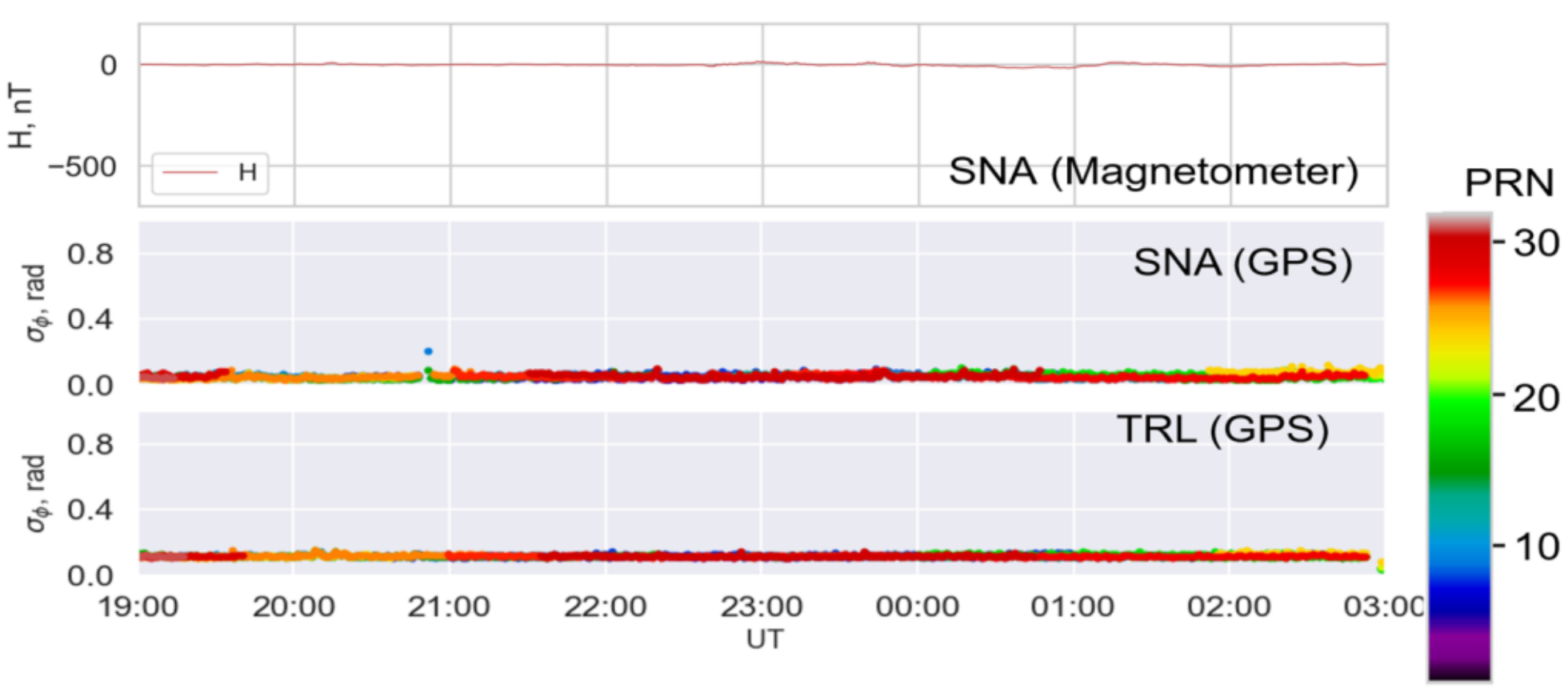

**Figure S1.** Variations in the H-component of the magnetic field measured at SANAE IV (SNA) on 9 May 2019 (mean field removed) and the corresponding 1-minute phase scintillation index ($\sigma_\phi$) for GPS signals observed at SNA and Troll (TRL). Different colours denote different satellite PRNs.

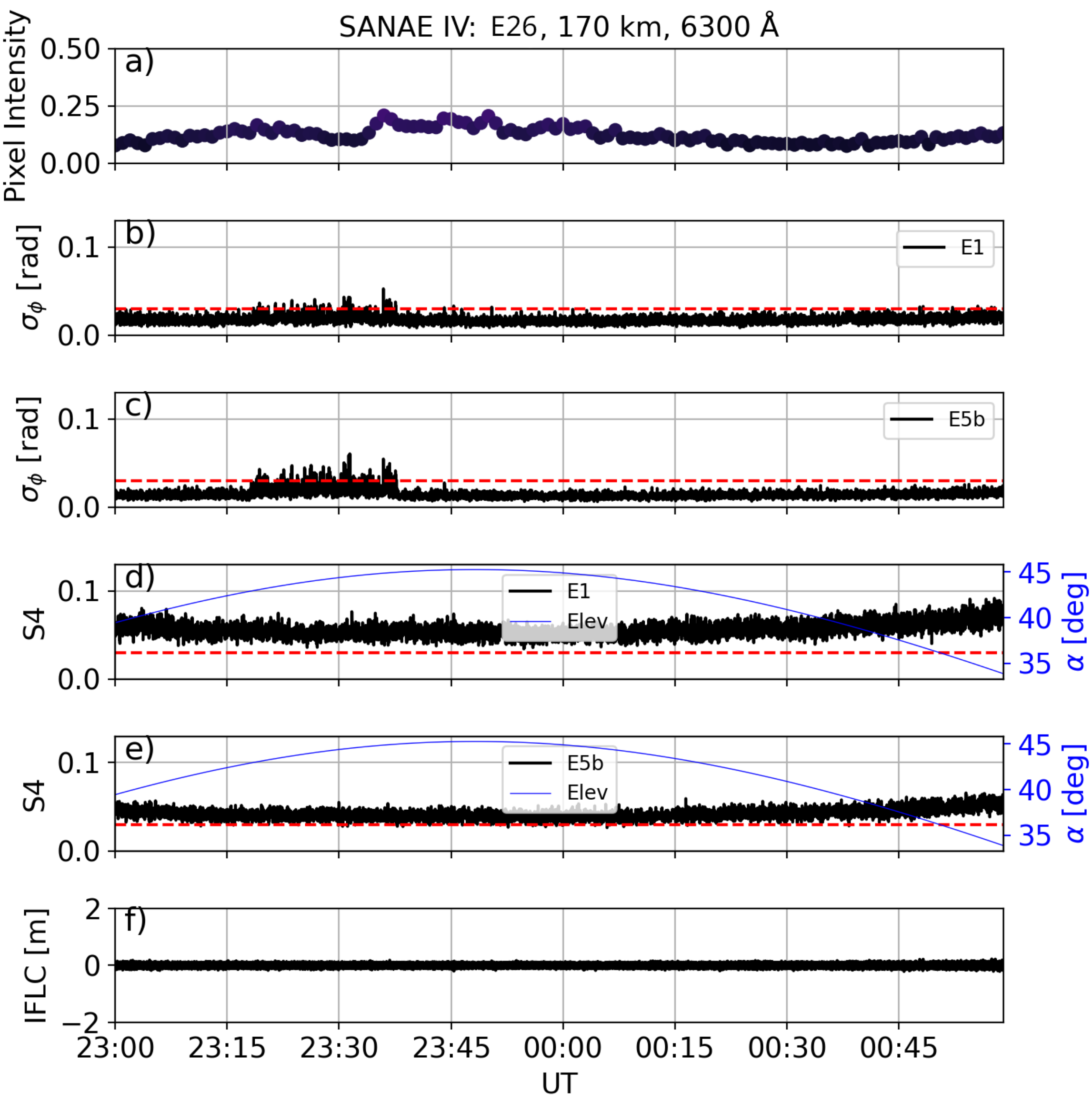

**Figure S2.** (a) Pixel intensity at the respective IPP position (170 km) of Galileo PRN E26 (96 in movie), 1-second $\sigma_\phi$ for (b) E1 and (c) E5a frequencies recorded by the Septentrio PolaRx5S receiver, 1-second *S*4 for (d) E1 and (e) E5a signals plotted together with the elevation angle in blue (axis in the right), (f) ionospheric free linear combination (IFLC). The red dotted line indicates the noise level of the receiver.

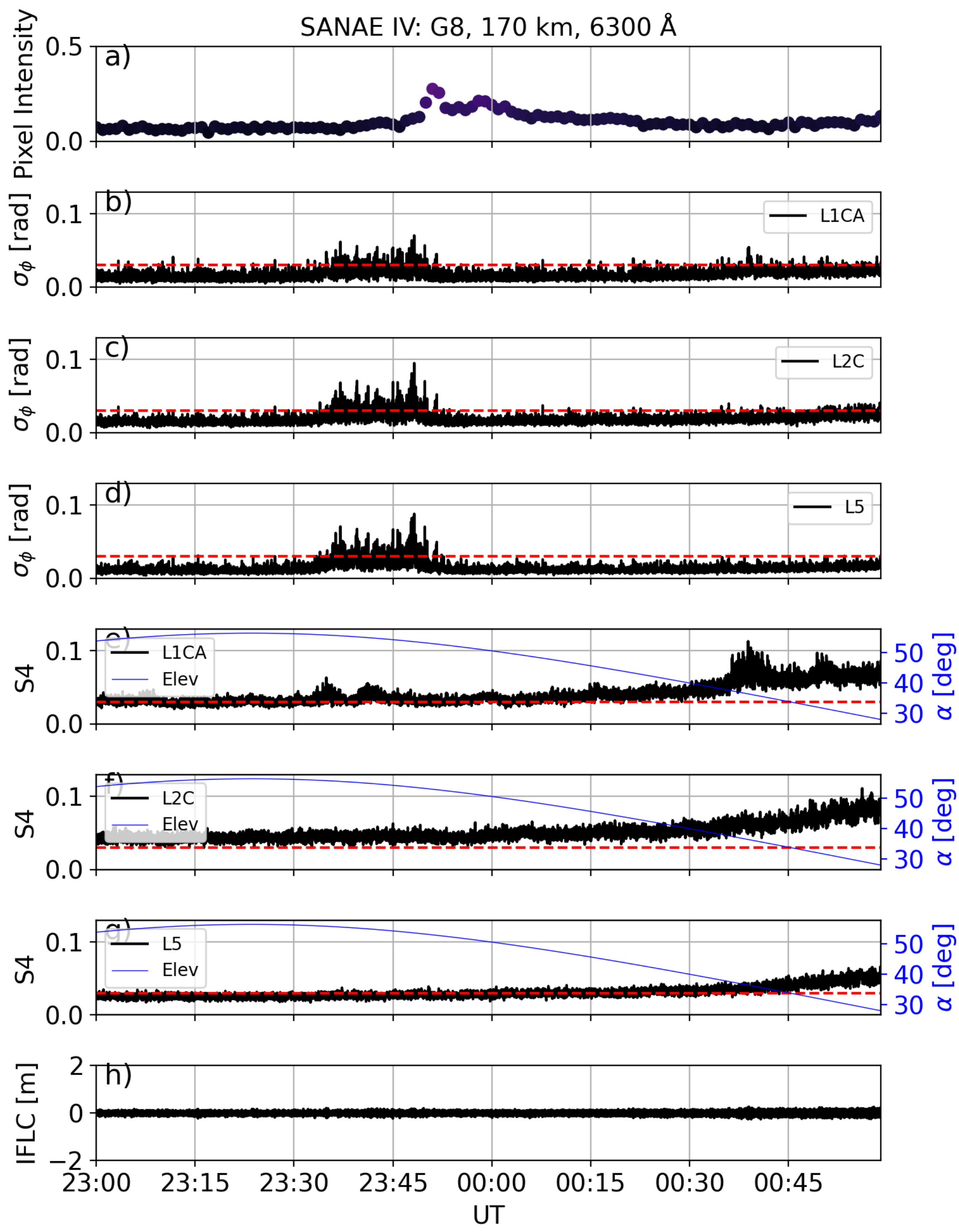

**Figure S3.** (a) Pixel intensity at the respective IPP position (170 km) of GPS PRN G08 (8), 1-second $\sigma_\phi$ for (b) L1CA, (c) L2C and (d) L5 frequencies recorded by the Septentrio PolaRx5S receiver, 1-second *S4* for (e) L1CA, (f) L2C and (g) L5 signals plotted together with the elevation angle in blue (axis in the right), (h) ionospheric free linear combination (IFLC).

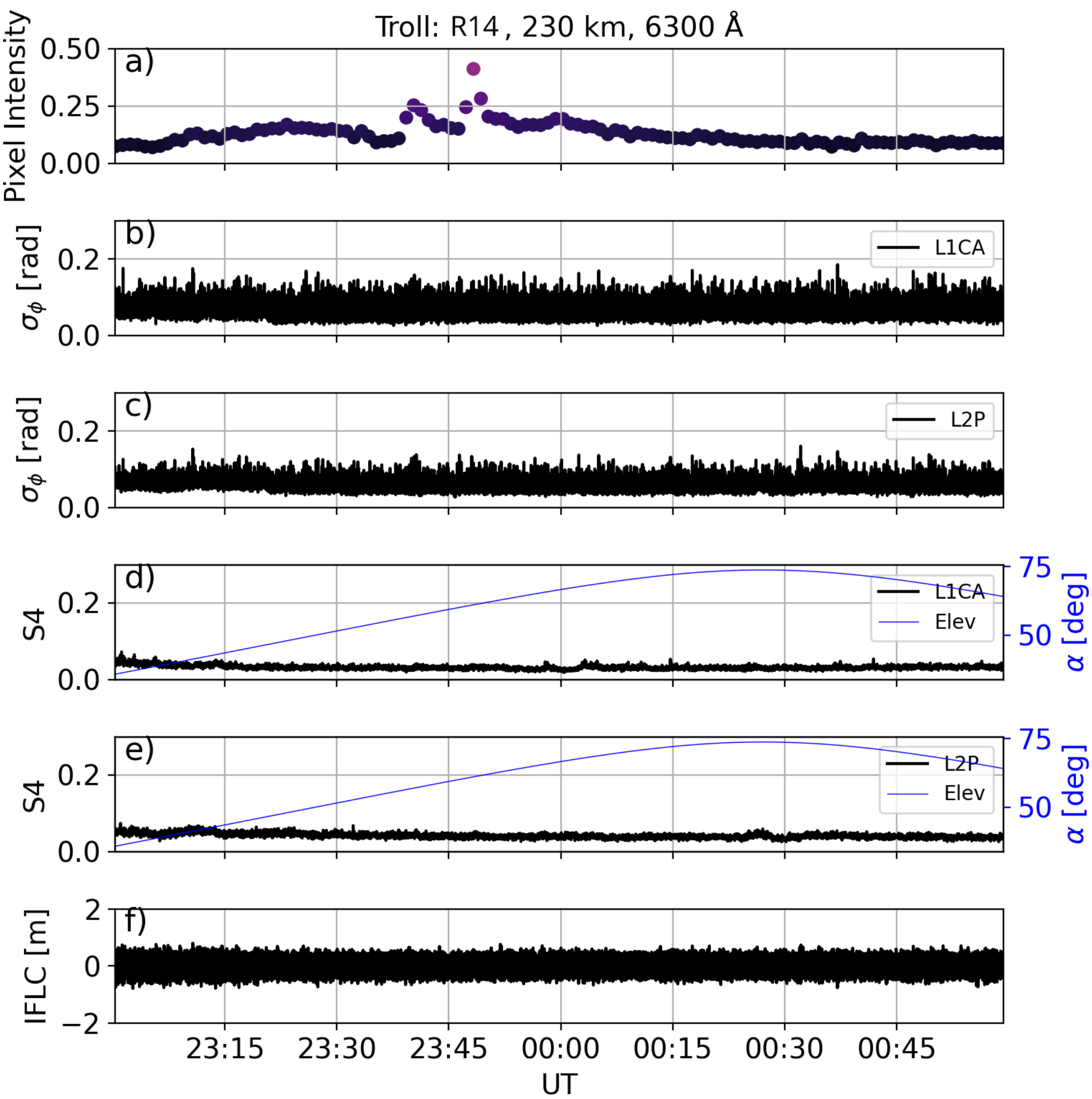

**Figure S4.** (a) Pixel intensity at the respective IPP position (230 km) of GLONASS PRN R14 (54), 1-second $\sigma_\phi$ for (b) L1CA and (c) L2P frequencies recorded by the NovAtel GPStation-6 receiver, 1-second $S4$ for (d) L1CA and (e) L2P signals plotted together with elevation angle in blue (axis in the right), (f) ionospheric free linear combination (IFLC).

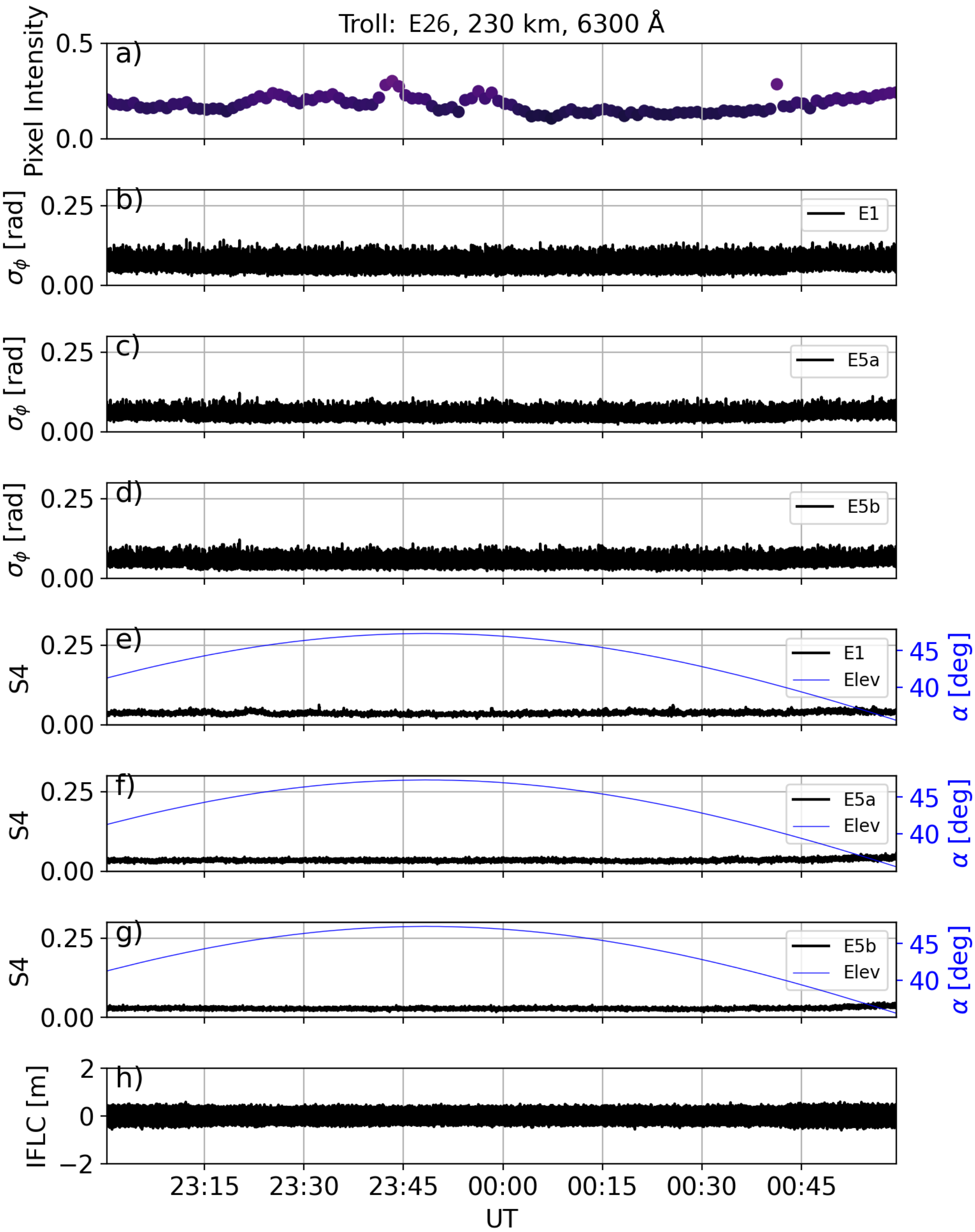

**Figure S5.** An example of satellite that passed through the discrete aurora or picket fence and did not experience fluctuations in phase and amplitude measurements. (a) Pixel intensity at the respective IPP position of the Galileo satellite PRN E26 (96), scintillation indexes $\sigma_\phi$ and S4 were calculated for E1 (b and e), E5a (c and f), and E5b (d and g) signals and were recorded by the NovAtel GPStation-6 receiver at the Troll station. S4 indexes (e-g) plotted together with the elevation angle in blue (axis in the right), (f) ionospheric free linear combination (IFLC).

Figure S6 shows SuperDARN convection patterns during the STEVE event. SANAE IV (SAN on maps), located within the region where STEVE was observed, remains near the boundary of convection cells in the evening sector (around 21:30 MLT). The overall convection pattern is relatively stable, consistent with the generally quiet geomagnetic conditions indicated by the low Kp and SYM-H indices. No major reconfiguration of the large-scale convection system is apparent during the interval of STEVE observations. Around 23:30 UT, the cross-polar cap potential appears to increase substantially (from ~35-37 kV to ~62 kV), and the flow vectors near midnight become stronger. That suggests localised electrodynamic activity despite otherwise quiet large-scale geomagnetic conditions.

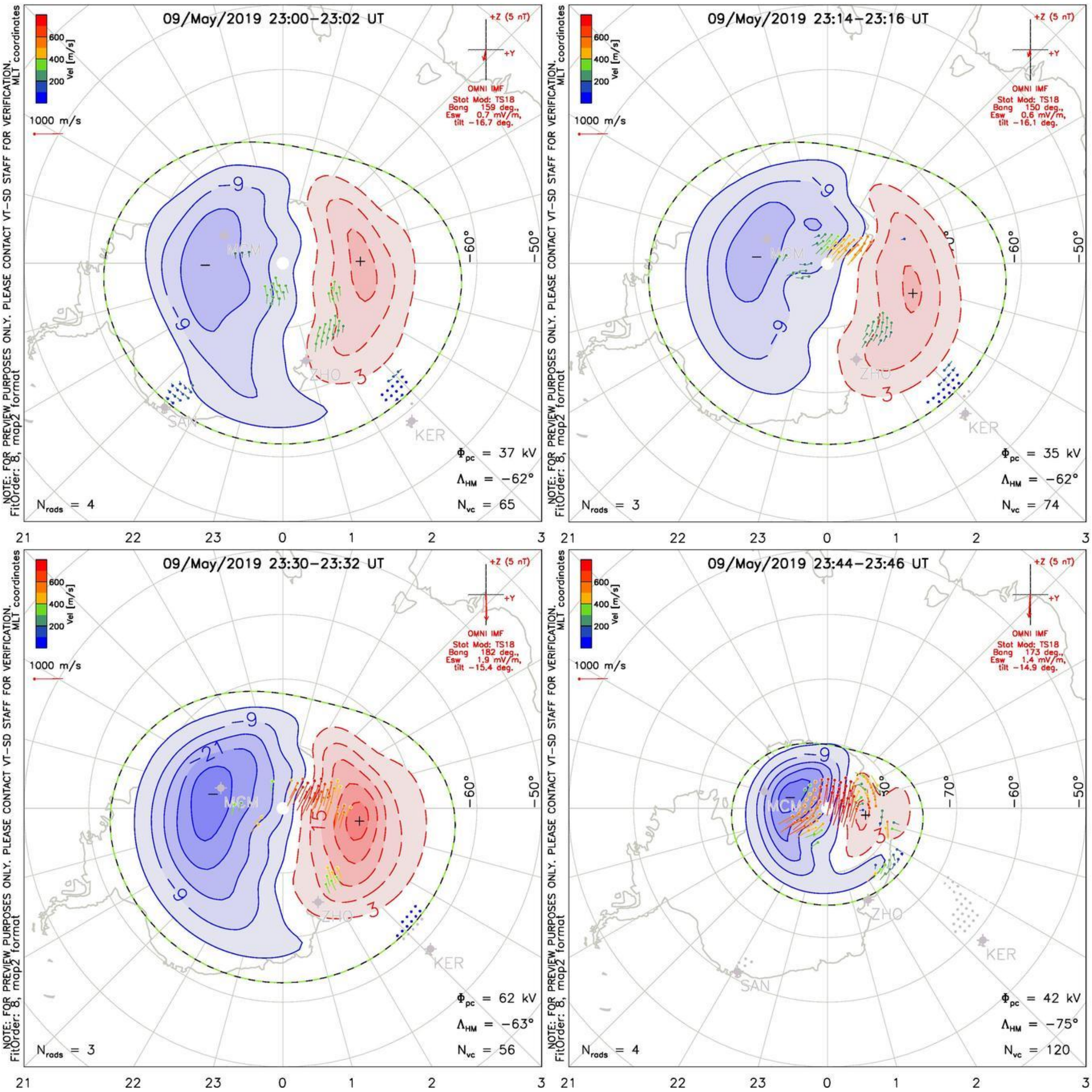

**Figure S6.** SuperDARN convection maps from Virginia Tech for 23:00, 23:15: 23:30, and 23:45 UT. The SuperDARN maps indicate that STEVE developed during a period of relatively modest but organised large-scale plasma convection, emphasising that detectable GNSS perturbations occurred under conditions that would generally be considered geomagnetically quiet.

**Problem encountered with LEO satellite data.**

To complete the picture of the background upper atmosphere conditions, we investigated data from low-Earth Orbit (LEO) satellites. During the observation period of STEVE, there was only one flyby of the European Space Agency satellite Swarm B (Wood et al., 2022 and references therein) over the Dronning Maud land, Antarctica. Unfortunately, the flyby data are contaminated (in terms of electron density, temperature, and ion flow velocity), see Figure S7. The electron density values are very low and around 23:53:27 UT they begin to fall from about 5300 $cm^{-3}$ to 500 $cm^{-3}$. This time is connected with increased measurements of electron temperature. The calibration flag bit at most in the cross-track ion flow dataset (EFIxTCT) indicates that the 16 Hz ion flow magnitude exceeded 8 km/s. Notably, errors are also contained in other flybys over the southern hemisphere, while in the northern hemisphere, there are no erroneous observations.

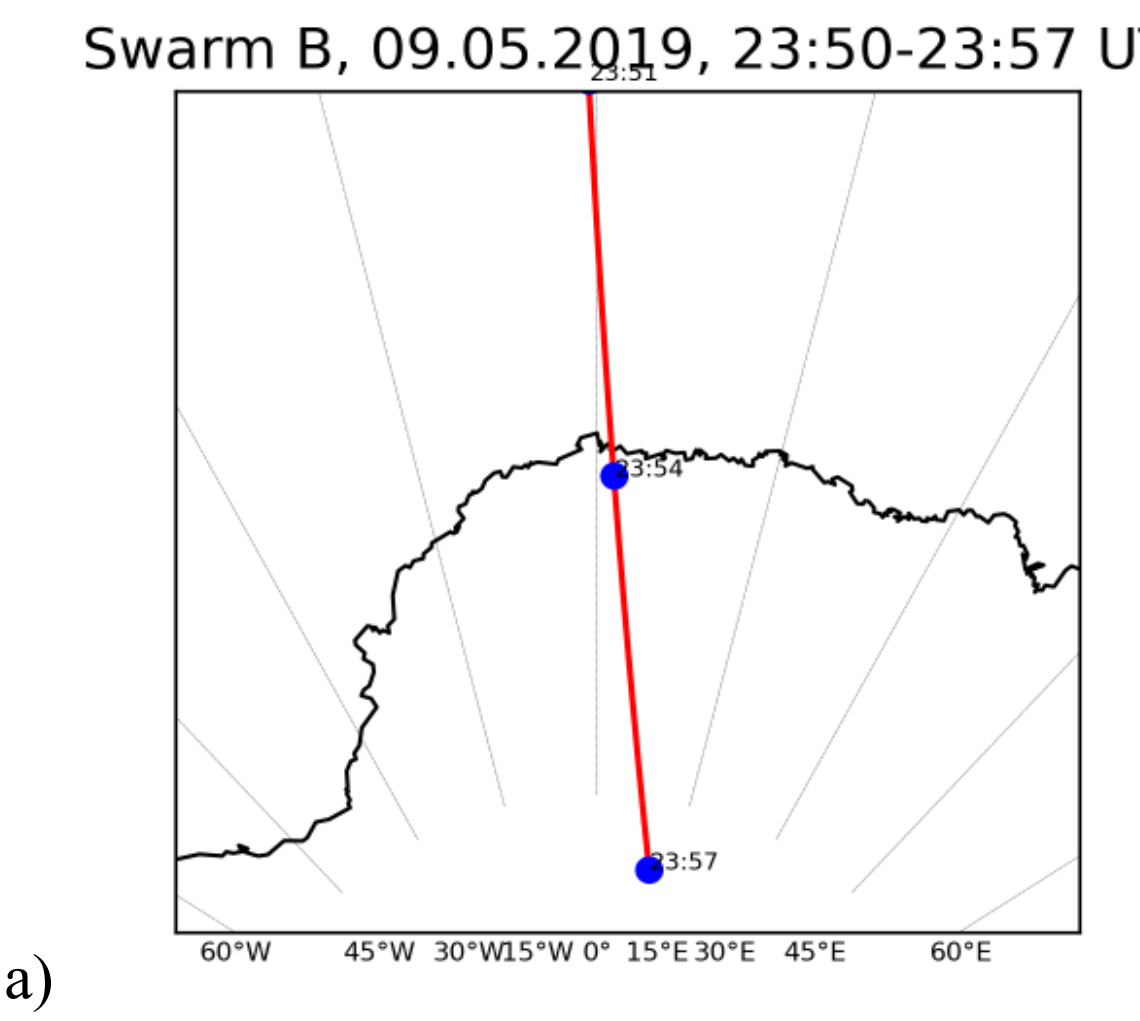

a)

b)

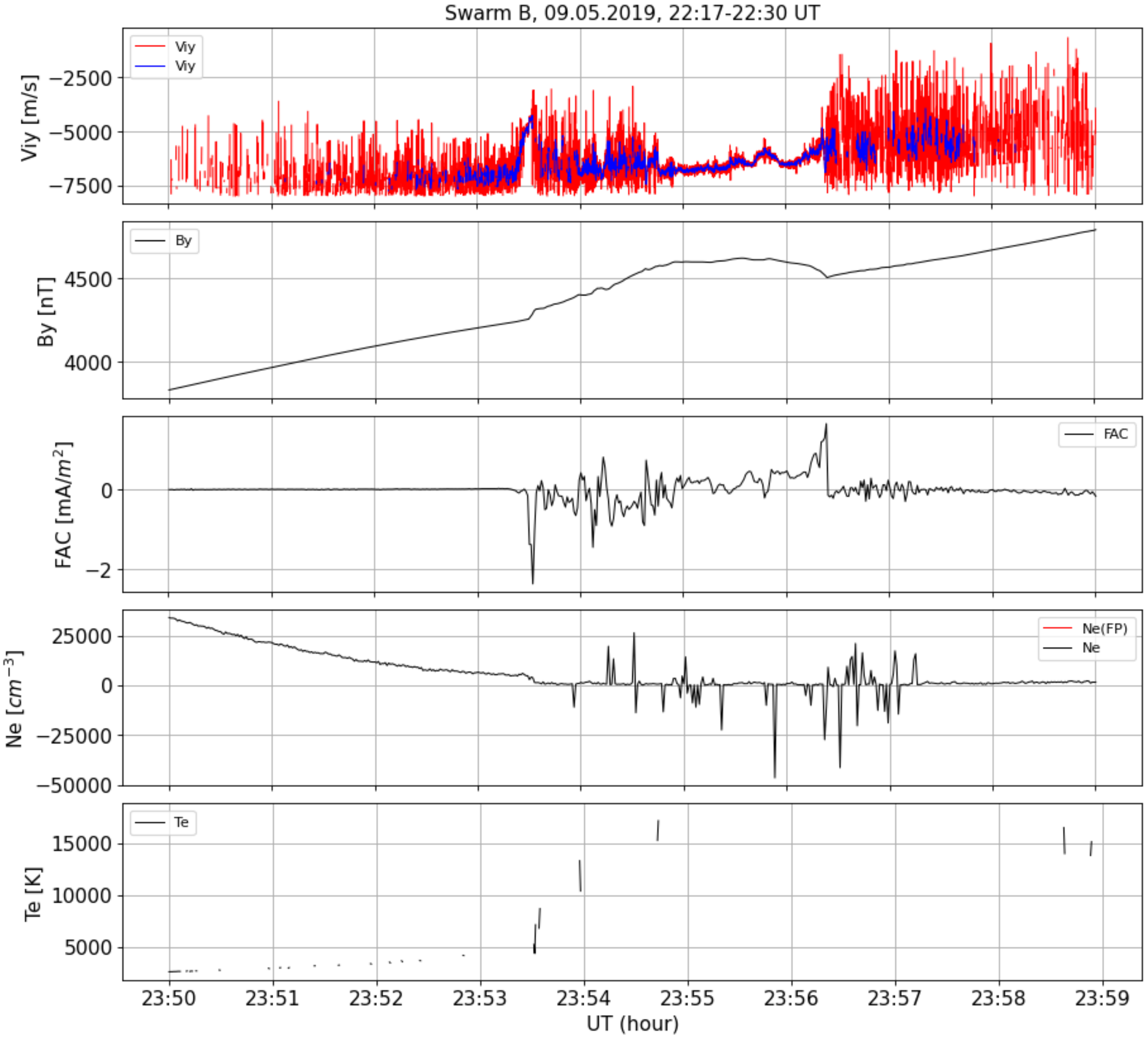

**Figure S7.** Swarm B satellite path over the Dronning Mound land (a) during STEVE event observed on May 9, 2019, and measured parameters (b) from top panel to bottom: ion velocity (Viy), magnetic field By component, vertical FAC, electron density (Ne) and temperature (Te). Unprocessed electron density data are shown. No faceplate measurements were available for this interval.

A similar problem (no density and temperature data, as well as drift velocities) accompanied the data from the DMSP satellites for the STEVE event (Fig. S8). The most suitable flight time (however, more in the direction from east to west and poleward of the station) was for the F18 satellite at about 23:23 UT and for the F16 satellite at about 23:47 UT. Therefore, there is no clean in situ satellite data based on which we could obtain information about the drift velocity or the thickness or location of SAID. Thus, the main data of the findings of this study are the data of scintillation receivers, the all-sky camera, and the all-sky imager.

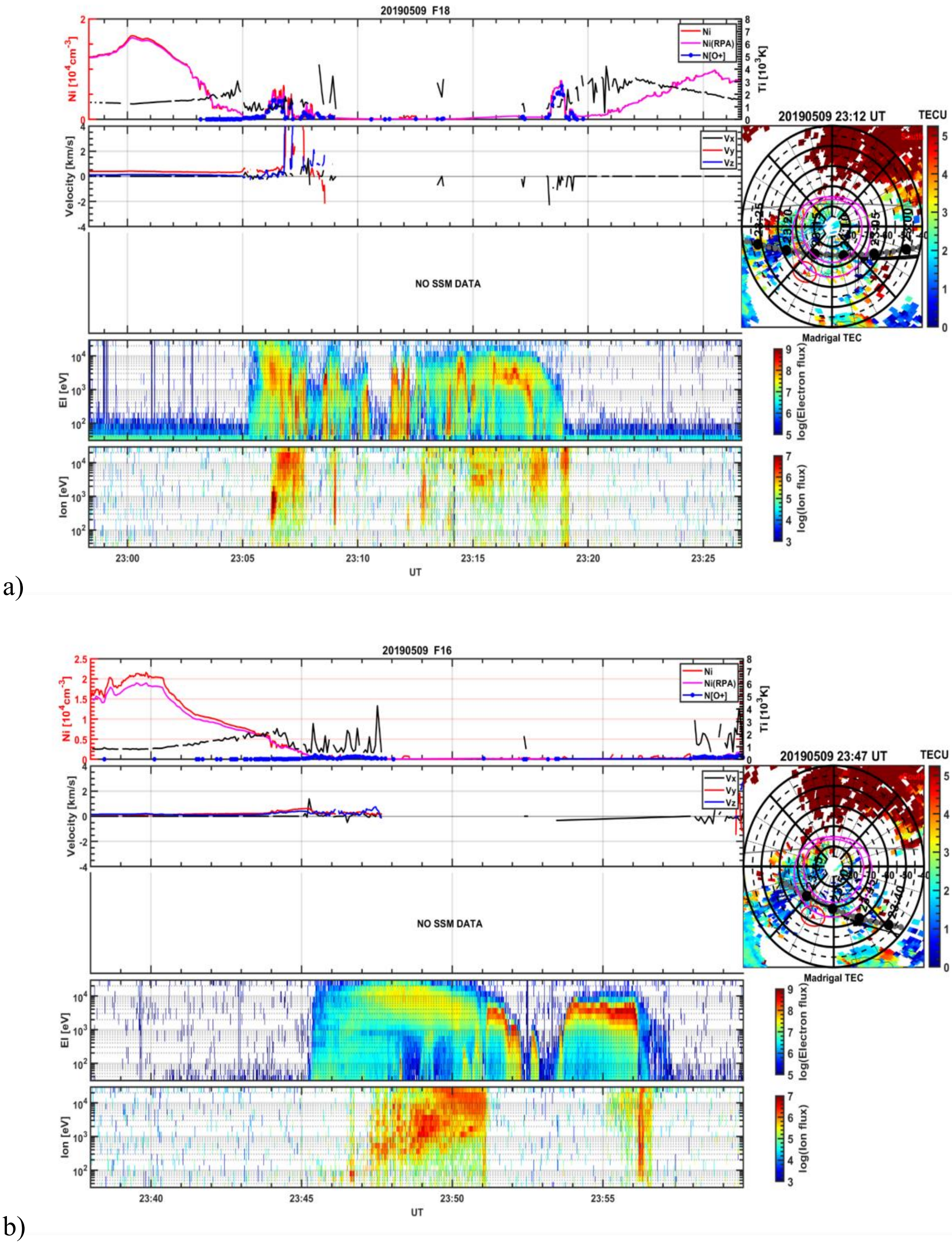

**Figure S8.** DMSP F18 (a) and F16 (b) observations during the STEVE event. From top to bottom, the panels show ion density (colour-coded according to the legend) and ion temperature (black line, right axis), ion velocity components, and electron and ion energy fluxes. The maps on the right panels show the total electron content (TEC) from the Madrigal database in magnetic coordinates, the DMSP satellite's trajectory over the region by black line, and the corresponding time. The solid grey line shows the solar terminator position. The statistical Feldstein auroral oval is indicated in purple, and the Troll station and its field of view are marked by a red triangle and red circle, respectively.

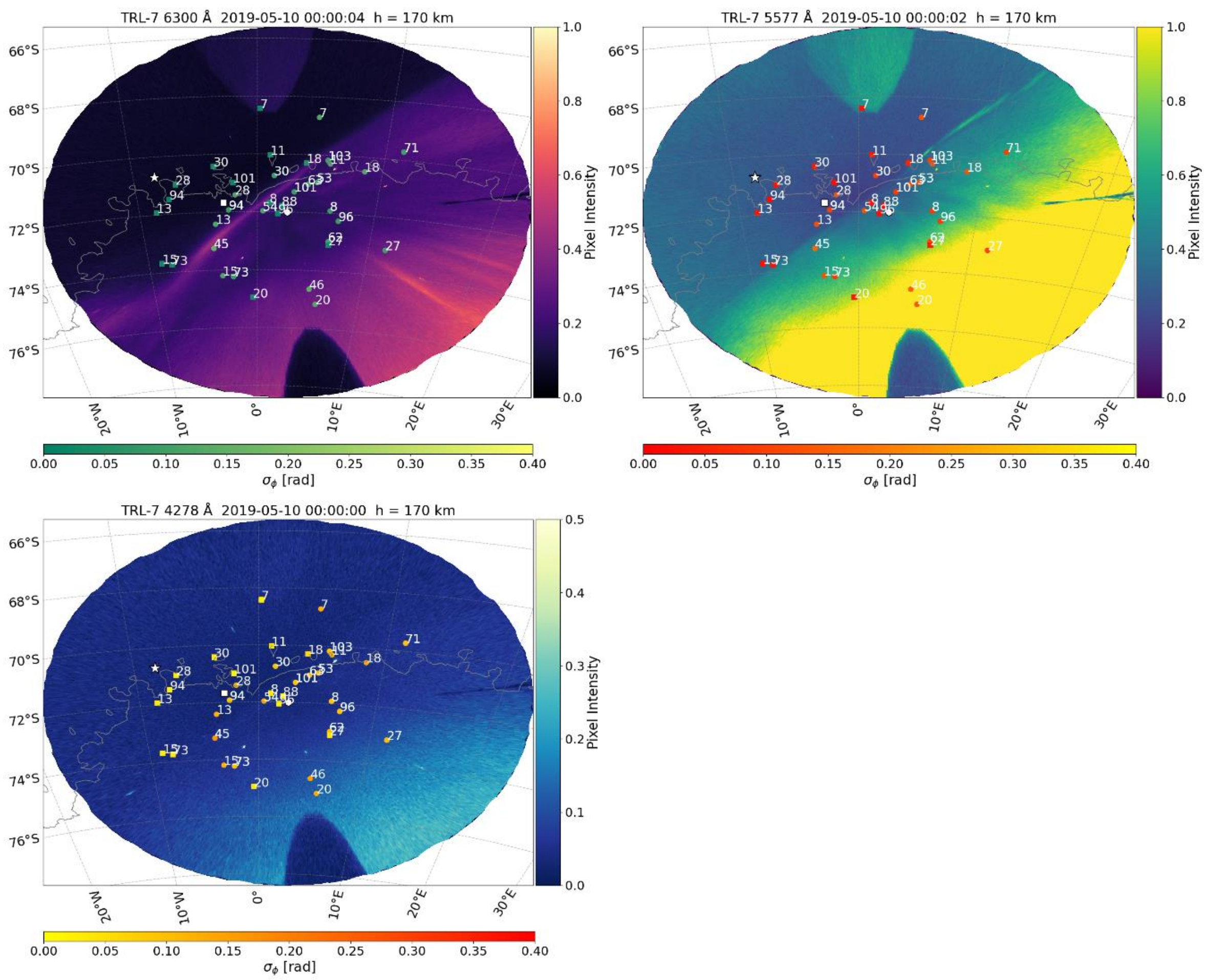

**Figure S9.** Comparison of STEVE observations obtained with different ASI filters. The 630.0 nm emission provides the clearest representation of the STEVE arc, while weaker structures are visible at 557.7 nm and are largely absent at 427.8 nm.